\documentclass{aa}
\usepackage{graphicx}
\usepackage{txfonts}
\usepackage[figuresright]{rotating}
\newcommand{\flux}{erg cm$^{-2}$ s$^{-1}$}

\begin{document}
   \title{XMM-{\em Newton} observations of the Lockman Hole: X-ray source catalogue and number 
counts\thanks{Based on observations obtained with XMM-{\em Newton}, an ESA science mission with 
instruments and contributions directly funded by ESA Member States and NASA}}

   \author{H. Brunner\inst{1}, N. Cappelluti\inst{1}, G. Hasinger\inst{1}, 
           X. Barcons\inst{2}, A.~C. Fabian\inst{3},  
           V. Mainieri\inst{1,4} \and G. Szokoly\inst{1}
          }

   \titlerunning{XMM-{\em Newton} observations of the Lockman Hole}

   \authorrunning{H. Brunner et al.}

   \offprints{H. Brunner, \email{hbrunner@mpe.mpg.de}
              }

   \institute{Max-Planck-Institut f\"ur extraterrestrische Physik,
              85478 Garching, Germany
              \and
              Instituto de F\'{i}sica de Cantabria (CSIC-UC), 39005 Santander, Spain
             \and
	     Institute of Astronomy, University of Cambridge, Madingley Road,
		Cambridge, CB3 0HA	
		\and
	     ESO,Karl-Schwarschild-Strasse 2, 85748 Garching, Germany	
			}

   \date{Received ; accepted}

   \abstract
{The Lockman Hole field represents the sky area of lowest Galactic line-of-sight 
column density N$_{\rm H}$=5.7$\times$10$^{19}$ cm$^{-2}$. It was observed by the XMM-{\em Newton} X-ray observatory 
in 18 pointings for a total of 1.16 Msec (raw EPIC pn observing time)  
constituting the deepest XMM-{\em Newton} exposure so far. After the removal of time intervals 
with high particle background, the final effective exposure was 637 ks.}
{We present a catalogue of the  X-ray sources detected in the central 0.196 deg$^{2}$ of the field and 
discuss the derived number counts and X-ray colours.}
{The analysis was performed using  the XMM-SAS data 
analysis package version 6.0. The detection sensitivity and fraction of spurious detections
was calibrated using Monte Carlo simulations.}
{In the 0.5--2.0~keV band, a sensitivity limit (defined as the faintest detectable source)
of 1.9$\times10^{-16}$ erg cm$^{-2}$ s$^{-1}$ was reached. 
The 2.0--10.0~keV band and 5.0--10.0~keV band sensitivity limits
were 9$\times10^{-16}$ erg cm$^{-2}$ s$^{-1}$ and
1.8$\times10^{-15}$ erg cm$^{-2}$ s$^{-1}$, respectively.
A total of 409 sources above a detection likelihood of 10 (3.9 sigma)
were found within a radius of 15$^\prime$ off the field centre, of which 
340, 266, and 98 sources were detected in the soft,
hard, and very hard bands, respectively. The number counts in each energy band
are in close agreement with results from previous surveys and with the synthesis
models of the  X-ray background. A 6\%  of Compton-thick source candidates have been 
selected from the  X-ray colour-colour diagram. This fraction is consistent with the 
most recent predictions of  X-ray background population synthesis 
models at our flux limits.
We also estimated, for the first time, the logN-logS relation for Compton-thick AGN.  
}
{}
 

   \maketitle
%
\section{Introduction}

Due to its extremely low line of sight column density 
of N$_{\rm H}$=5.7$\times$10$^{19}$ cm$^{-2}$, the
Lockman Hole field is one of the main extragalactic deep-survey  
target areas of recent X-ray observatories.  
The choice of a low N$_{\rm H}$  field on the one hand optimizes the soft-band sensitivity
compared to fields with higher N$_{\rm H}$. An even more important aspect is that
for many sources we can determine spectra to the lowest possible energies and
thus get a better handle on the continuum and intrinsic absorption.

 The field was observed for a total of 1.1 Msec, by the ROSAT HRI detector, 
reaching a sensitivity limit of 
1.2~$\cdot 10^{-15}$ erg cm$^{-2}$ s$^{-1}$ in the 0.5--2.0~keV 
band, for the first time resolving the majority of the soft X-ray
background into discrete sources (Hasinger et al. \cite{hasinger98}). 
The Lockman Hole was also the target of
the deepest observations conducted by the XMM-{\em Newton}
observatory, considerably improving the sensitivity limit in the 
soft band and extending the survey into the hard X-ray band up to 
an energy of 10 keV. Although 
the sensitivity limit of the XMM-{\em Newton} Lockman Hole survey does not 
reach the  sensitivity of the deepest
surveys performed by the {\em Chandra} observatory (i.e. the {\em  Chandra} deep 
field North and South, Bauer et al. \cite{bauer}, Giacconi et al. \cite{giac}), the larger
collecting area of XMM-{\em Newton} offers the possibility to perform
a spectral analysis of a large fraction of the 
detected sources down to the confusion limit, opening up a 
powerful window into the cosmological evolution of the X-ray 
universe.  

 The Lockman Hole field dataset allowed us for example to study the Fe K line emission of one of the most 
distant X-ray selected cluster of galaxies, RX J1053.7+5735l.
Hashimoto et al. (\cite{hashimoto}) were able to determine the X-ray 
redshift of the cluster and its 
metallicity. The XMM-{\em Newton} Lockman Hole survey is also well suited
to investigate the X-ray spectra and cosmological evolution of 
both absorbed and unabsorbed AGN.  With its  large  collecting area,  XMM-{\em Newton}
is able to obtain detailed spectral information even for the faintest  
sources.
Mainieri et al. (\cite{mainieri}) and Mateos et al. (\cite{mateos}) 
performed a detailed spectral analysis
and they discovered several candidate Compton thick AGNs. 
According to the recent model of Gilli et al. (2007),  40\% of the flux  of
the  X-ray background is expected to arise in  highly  absorbed 
AGNs (N$_{\rm H}>$10$^{22}$ cm$^{-2}$). These sources are mostly faint
and detectable with deep X-ray surveys  such as the XMM-{\em Newton} Lockman Hole  survey. 
Worsley et al. (\cite{worsley04}) estimated that  the XMM-{\em Newton} Lockman Hole observations
resolve $\sim$100\% of the soft  X-ray background into discrete sources while at  energies above 5 keV 
this fraction drops to $\sim$50\%. The missing part therefore could be  
due to very faint, highly  absorbed AGNs (N$_{\rm H}>$10$^{22}$ cm$^{-2}$) to which  
XMM-{\em Newton} is not sensitive.

\section{Observations}
\begin{figure*}
\centering
\includegraphics[angle=-90,width=17cm,clip]{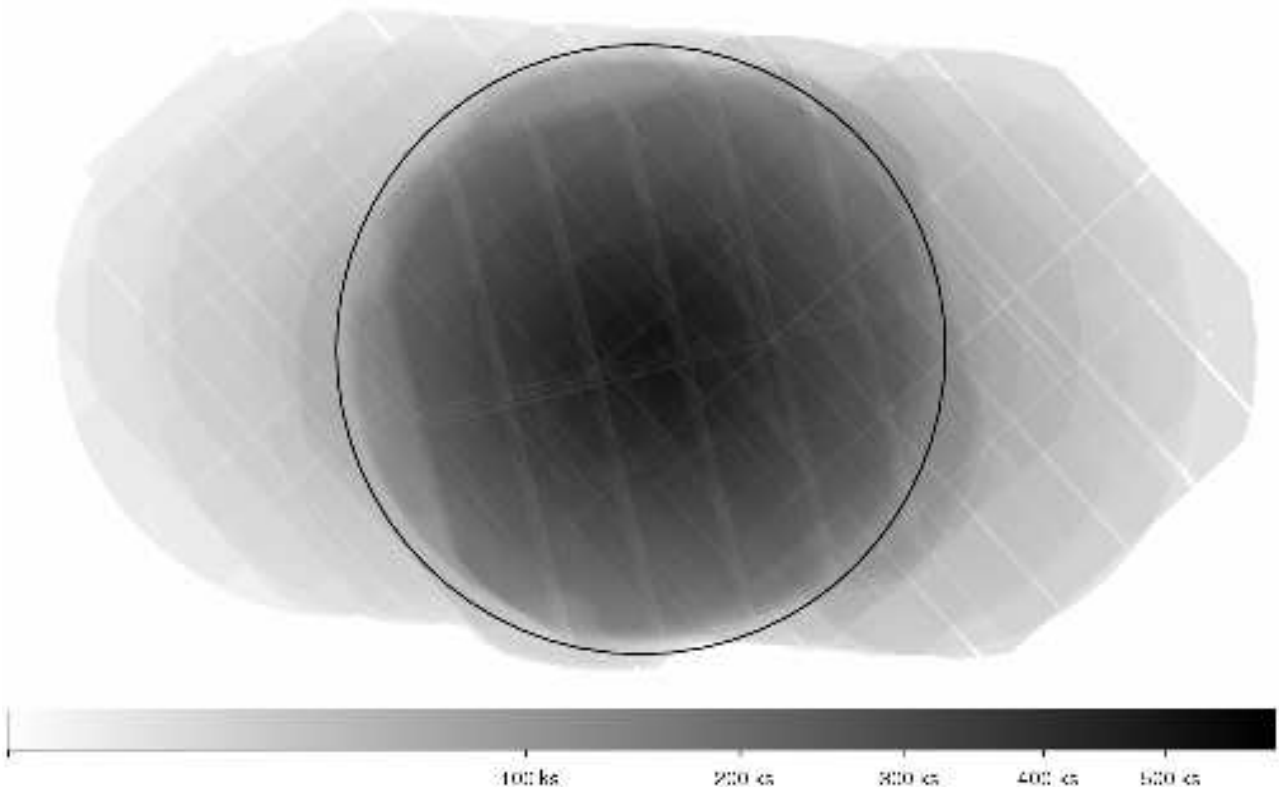}
\caption{Combined all-EPIC vignetting-corrected soft band 
exposure map of 18 XMM-{\em Newton} pointings on the Lockman Hole field. The black circle represents the region of 15$^\prime$ radius under investigation.}
\label{LHexp}
\end{figure*}
The Lockman Hole field was observed by XMM-{\em Newton} in 18 individual
pointings distributed over a two and a half year time period 
from April 2000 to December 2002. Twelve  pointings were centred
at the approximate field centre coordinates +10$^h$ 52$^{m}$ 43$^{s}$, 
57$^\circ$ 28$^\prime$ 48$^{\prime\prime}$ (J2000),
with small offsets $\sim$15$^{\prime\prime}$ to  achieve an adequate coverage of the
sky area falling into chip gaps in individual observations.
The remaining  pointings were spread out over about 30$^\prime$ in right
ascension, thus covering a somewhat wider area at intermediate exposure levels.
The pattern of the observations is shown in the vignetting corrected 
exposure map of Fig. \ref{LHexp} (see next section). 
The total XMM-{\em Newton} exposure time spent on the field was
1.30 Msec (EPIC MOS detectors; EPIC pn detector: 1.16 Msec). 
After removal of times affected by background flares
765 ks (EPIC MOS detector; EPIC pn detector: 637 ksec) are left which 
are suitable for  scientific analysis.
If the loss of sensitivity at large off-axis angles due to vignetting
is considered, the average exposure time is
reduced to 557.6 ks (weighted average of MOS and pn detectors). 
This paper addresses the sky area within 15$^\prime$ of the field centre
where the exposure exceeds 185 ks. 
Table~\ref{lhObs} lists the pointing directions and exposure times.
The twelve central pointings and  East and  West offset pointings are
marked by C,  E, and  W, respectively.

\begin{table*}
\caption[]{Pointing directions and exposure times}
\label{lhObs}
\begin{center}
\begin{tabular}{lllllcll}
\hline
rev.$^{a}$&OBS\_ID$^{b}$&&\multicolumn{2}{c}{RA~~~(J2000)~~~DEC} & offset$^c$&
                                            \multicolumn{1}{c}{MOS} & \multicolumn{1}{c}{pn} \\
      &           && hh mm ss & $^\circ$~~~$^\prime$~~~$^{\prime\prime}$ &$^\prime$&
                                               \multicolumn{2}{c}{exposure times$^d$ [ksec]} \\
\hline
070 & 0123700101 &C$^e$& 10 52 41.0 & 57 27 07 & 1.7 & 43.9 (33.7)  & 46.6 (33.6) \\
071 & 0123700201 &C    & 10 52 42 4 & 57 27 16 & 1.5 & 61.0 (38.0)  & 55.9 (31.8) \\
073 & 0123700401 &C    & 10 52 41.7 & 57 27 06 & 1.7 & 15.0 (12.9)  & 16.4 (13.2) \\
074 & 0123700901 &C    & 10 52 41.1 & 57 27 17 & 1.5 & 15.0 (4.5)   & 14.0 (4.5) \\
081 & 0123701001 &C    & 10 52 40.1 & 57 27 19 & 1.7 & 36.4 (25.7)  & 37.2 (25.6) \\
344 & 0022740101 &C    & 10 52 45.6 & 57 30 27 & 1.8 & 83.3 (0.3)   & 75.3 (0.0) \\
345 & 0022740201 &C    & 10 52 46.1 & 57 30 28 & 1.7 &63.9 (23.6)  & 61.2 (26.9) \\
349 & 0022740301 &C    & 10 52 43.9 & 57 30 27 & 1.7 & 37.7 (33.5)  & 36.4 (29.0) \\
522 & 0147510101 &  W    & 10 51 05.6 & 57 29 33 &13.1 &  92.6 (69.8)  & 91.0 (52.5) \\
523 & 0147510801 & W    & 10 51 29.8 & 57 29 50 & 9.9 &  77.2 (39.4)  & 75.7 (18.7) \\
524 & 0147510901 &C    & 10 52 44.8 & 57 30 26 & 1.7 & 90.2 (40.6)  & 88.5 (18.9) \\
525 & 0147511001 & W    & 10 52 10.4 & 57 30 15 & 4.6 & 83.4 (75.7)  & 81.9 (63.1) \\
526 & 0147511101 & E    & 10 53 20.3 & 57 30 52 & 5.4 & 97.2 (34.9)  & 93.4 (24.2) \\
527 & 0147511201 & E    & 10 54 00.4 & 57 31 09 &10.7 &  101.5 (28.2) &100.5 (21.9) \\
528 & 0147511301 & E    & 10 54 31.9 & 57 31 30 &14.9 & 85.0 (22.2)  & 82.5 (15.2) \\
544 & 0147511601 &C    & 10 52 38.7 & 57 30 26 & 2.0 &  124.9 (102.7)&121.1 (95.8) \\
547 & 0147511701 &C    & 10 52 40.6 & 57 28 29 & 1.4 &  100.4 (96.5) & 98.8 (91.4) \\
548 & 0147511901 &C    & 10 52 41.0 & 57 30 45 & 2.0 &   90.4 (82.9)  & 88.8 (71.0) \\
\hline
\end{tabular}
\begin{list}{}{}
\item[$^a$] XMM-{\em Newton} orbit number (revolution)
\item[$^b$] XMM-{\em Newton} observation ID
\item[$^c$] offset from field centre +10$^h$ 52$^{m}$ 43$^{s}$, 57$^\circ$ 28$^\prime$ 
48$^{\prime\prime}$
\item[$^d$] raw (after flare screening)
\item[$^e$]  central (C) pointings, East (E), and West (W) offset pointings
\end{list}
\end{center}
\end{table*}


\begin{figure*}
\centering
\includegraphics[scale=0.9]{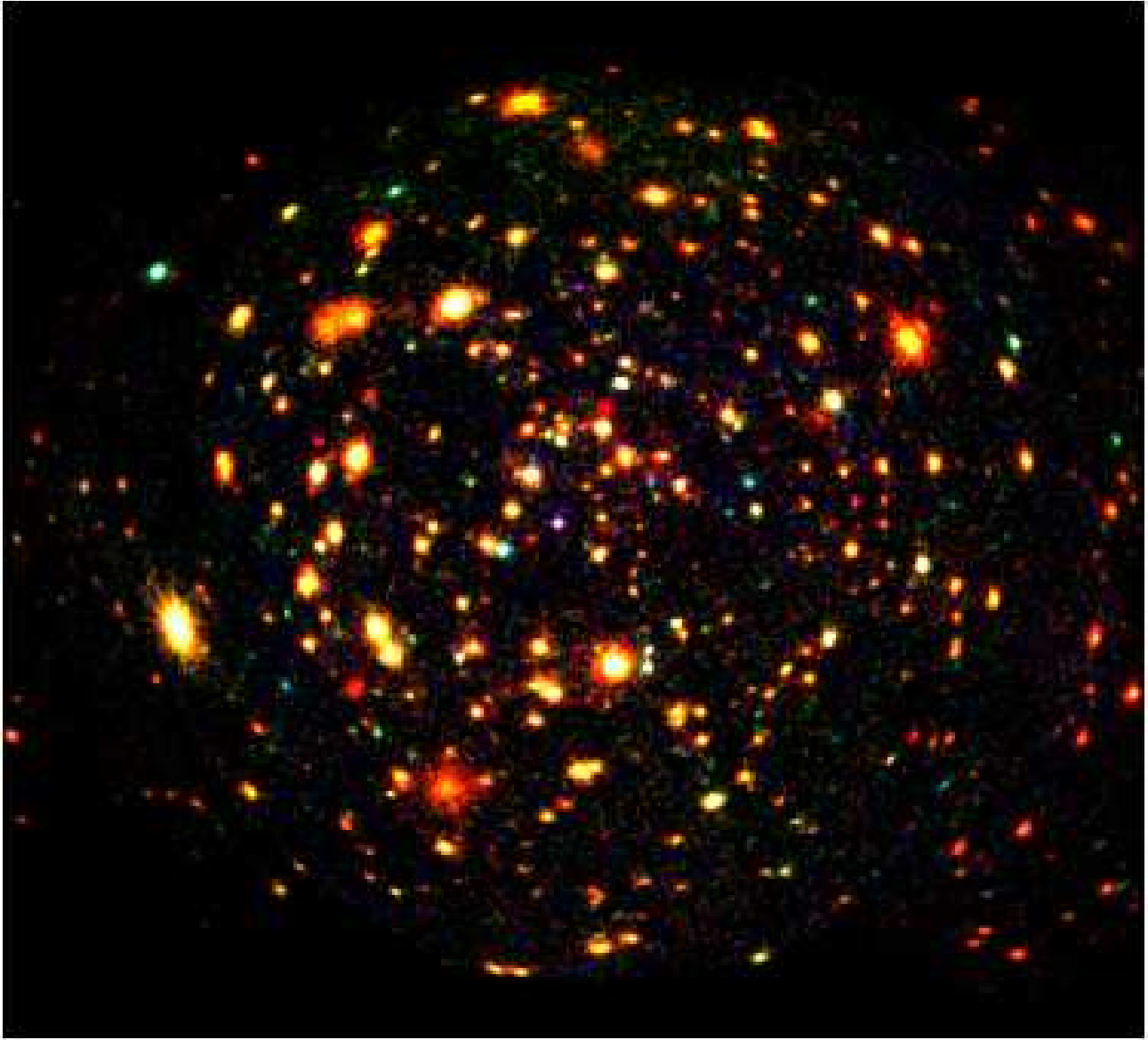}
\caption{Exposure corrected and background subtracted false colour image.
The colours red, green, and blue code the 0.5--2.0~keV, 2.0--4.5~keV, and 4.5--10~keV fluxes, 
respectively. 
North is at the top, East is to the left. The image is centred on 
10$^h$ 52$^m$ 43$^s$, 57$^\circ$ 28$^\prime$ 48$^{\prime\prime}$ (J2000) and the 
field size is approximately  30$^\prime$ $\times$ 30$^\prime$.}
\label{LHimgcorr}
\end{figure*}

\begin{figure*}
\centering
\includegraphics[angle=-90,width=17cm,angle=0,clip]{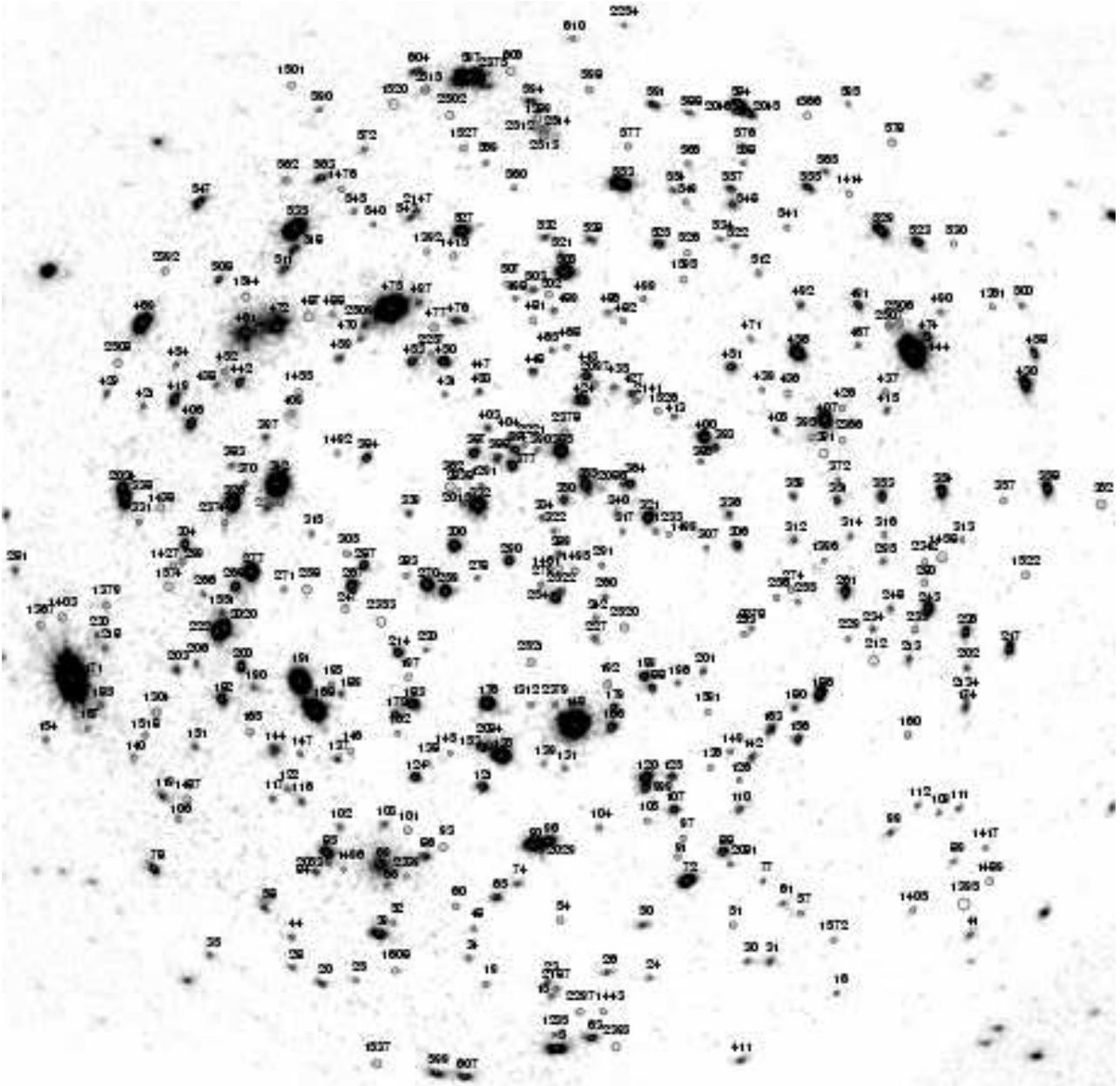}
\caption{ Background subtracted full band image. 
Detected sources are marked by their 1$\sigma$
positional error circle and source number.
Field size and orientation as in Fig. 2.}
\label{LHimg}
\end{figure*}

\section{Data reduction}

Event calibration was performed by uniformly processing all 18 data sets using
the XMM-SAS 6.0 software package\footnote{Available from the 
XMM-{\em Newton} Science Operation Centre at http://xmm.vilspa.esa.es/.}. 
Times affected by background flares
were screened out by using a 3$\sigma$ clipping method, 
resulting in the clean exposure times listed in 
table~\ref{lhObs}.  Out-of-time event\footnote{ X-ray events 
hitting the detector during read-out resulting in a loss of positional information
in the read-out direction.} files were  processed in the same way 
as  the data and subtracted from our observations.  
In order to achieve the full sensitivity,
data from all three XMM-{\em Newton} cameras of all observations were co-added on 
a joint sky-pixel grid of 3$^{\prime\prime}$ pixels after performing astrometric
corrections of each individual dataset based on highly accurate source 
positions from ROSAT optical follow-up work. 
The standard XMM-SAS source detection  package 
was used to perform source detection on the area within 15$^\prime$
from the field centre of the co-added data in the energy bands 0.5--2.0~keV,
2.0--4.5~keV and 4.5--10.0~keV. The spectral range between 7.8 and 8.2~keV, affected 
by instrumental Cu line background emission, was excluded from
the EPIC pn camera 4.5--10~keV images  to improve the detection sensitivity 
in that band. 

Unless otherwise stated  we analyzed the data in the three  non-overlapping energy bands
0.5--2.0~keV, 2.0--4.5~keV, and 4.5--10.0~keV. The fluxes measured in these energy bands were
extrapolated to the canonical 0.5--2~keV, 2--10~keV, and 5--10~keV bands, respectively,
 by assuming a power-law spectrum with spectral index $\Gamma=2$ and 
galactic N$_{\rm H}$=5.7 $\times$ 10$^{19}$ cm$^{-2}$.
For the sake of clarity we will call these energy bands soft, hard, and very hard band, 
respectively. The 0.5--10.0~keV energy band will be referred to as full band.

Exposure maps were computed in the same energy bands as the images.
Since this work has been conducted using co-added EPIC MOS and  pn data, the exposure maps 
were weighted and co-added according to the respective energy convertion factors, 
computed  from the  instrument response matrices, again assuming
a power-law spectral index of $\Gamma$ = 2.0 and N$_{\rm H}$=5.7 $\times$ 10$^{19}$ cm$^{-2}$.   
The combined all-EPIC exposure map
of the 18 observations of the Lockman Hole field is displayed
in Fig.~\ref{LHexp}. The displayed exposures include corrections for 
telescope vignetting (here displayed for the soft band).
The circle marks the 15$^\prime$ radius area 
discussed in this paper. 

The large variety of spectral properties of the source 
population is demonstrated in the false colour image of the field (Fig. \ref{LHimgcorr}), displaying
exposure corrected and background subtracted count rates of the 
soft, hard, and  very hard energy bands in the colours red, green, and blue,
respectively.  While the distribution of spectral  types  as indicated by the colour images 
appears wide,  Mateos et al. (\cite{mateos})  and  Mainieri et al. 
(\cite{mainieri} and \cite{mainieri2}) 
in the  Lockman Hole and Hasinger et al. (\cite{hasinger06}) in the COSMOS Field
showed that the  spectra of the X-ray sources have an average photon index $<\Gamma>\sim$2.
The choice of modeling the  average spectrum with a power-law with spectral index $<\Gamma>=$2 
and N$_{\rm H}$=5.7$\times$~10$^{-19}$ cm$^{-2}$ for estimating the fluxes 
is therefore consistent with a detailed spectral analysis. 

The considerable fraction of blue and green objects mainly
represents the population of absorbed AGN in the field. A quantitative
analysis of the X-ray colours based on the hardness ratios determined
from the source count rates in each energy band is presented in section
 4.4. Also note the clearly visible population of extended objects,
including the distant double cluster RX J1053.7+5735 
(Hashimoto et al. \cite{hashimoto} and \cite{hashimoto2}) 
in the upper left part of the image  (source numbers 461 and 472; Fig.~\ref{LHimg},
tables~\ref{souCat} and \ref{souCatExt}). The source parameters of the detected 
extended objects are also discussed in section  4.4.

\section{Data analysis}

\subsection{Maximum likelihood multi-stage source detection}

Source detection was performed simultaneously on the soft, hard, and
very hard band images using the {\em eboxdetect} and {\em emldetect} 
tasks of the XMM-SAS data analysis package. In the following section we
briefly describe the XMM-SAS source detection procedure; see  the
XMM-SAS documentation$^1$ for additional details. 

The source detection 
procedure consists of three consecutive detection steps. 
An initial source list is created by running a sliding box detection 
algorithm (XMM-SAS task {\em eboxdetect}) with detection box sizes 
$5\times5$, $10\times10$, and $20\times20$ pixels ($15^{\prime\prime}\times15^{\prime\prime}$, 
$30^{\prime\prime}\times30^{\prime\prime}$, and $60^{\prime\prime}\times60^{\prime\prime}$), 
where the local background is estimated from 
pixels adjacent to the detection box. The global background is 
determined by fitting a combination of vignetted and 
non-vignetted background components to the source free regions 
(as determined from the initial source list). 
The {\em eboxdetect} detection task is then rerun using the
global background model for improved detection sensitivity, creating
a list of source candidates down to a low statistical 
significance level ($\sim2\sigma$). This list is fed into the {\em emldetect}
task which performs a Maximum Likelihood fit  
of the distribution of source counts (based on the Cash C-statistics approach, Cash \cite{cash}),
using a point spread function model obtained from ray tracing calculations 
(medium accuracy model from the XMM-{\em Newton} calibration database), 
creating the final source list of best-fit source positions and fluxes.
The fit is performed simultaneously in all energy bands by summing the
likelihood contributions of each band. Sources exceeding the
detection likelihood threshold in the full band are regarded as detections;
the catalogue is thus full band selected, i.e., it also
includes very weak sources which do not exceed the detection 
threshold in any of the individual bands. 
Alternatively, it is possible that a weak source which slightly 
exceeds the detection threshold in one energy band and is not detected
in the other bands may be excluded from the catalogue because it does not
reach the detection threshold in the total band.
 A detection likelihood threshold of 10 was used for inclusion of detected objects
into the source catalogue while objects with detection likelihoods above 6 were 
considered for determining the number counts (see details in section 4.3 and 4.4).
The C-statistics approach can also be
used to calculate the likelihood of the source extent. The best-fit
parameters determined for each detected source thus are the source position, 
fluxes in each energy band and, optionally, the source extent 
(expressed as the core radius of a King profile). 
In the latter case, for each source, the distribution of source counts is modelled by 
numerically folding the  point spread function with a King profile. 
The significance of detection in the total band and in each individual 
energy band is expressed in the 
form of a detection likelihood, $\mathcal{L}$, normalized to two 
degrees of freedom, obeying the 
relationship $\mathcal{L}=-\ln(p)$, where $p$ is the probability that the
observed source counts are due to Poissonian fluctuations\footnote{Note, 
that due to the higher number of degrees of freedom in the total band
(one additional degree of freedom per band), the sum of 
the individual band detection likelihoods exceeds the total 
band detection likelihood by about 2.1.}. In principle the expected number 
of  spurious sources is thus defined as the number of independent 
 trials $\times~p$ with $p=e^{\mathcal{-L}}$. The number of trials can be 
approximated by the number of  point spread function sized beams in the field.
 This relationship is displayed as a straigth line in Fig.~\ref{spurDet} 
(see section 4.2 for details). However, 
due to the complex nature of the multi-step detection procedure, this 
relationship can only be regarded as a rough approximation which 
needs to be calibrated by Monte Carlo simulations (see section 4.2).
A detailed description of the Maximum Likelihood source detection
method can also be found in Cappelluti et al. (\cite{nico}).



\subsection{Monte Carlo simulations}

\begin{figure}
\resizebox{\hsize}{!}{\includegraphics{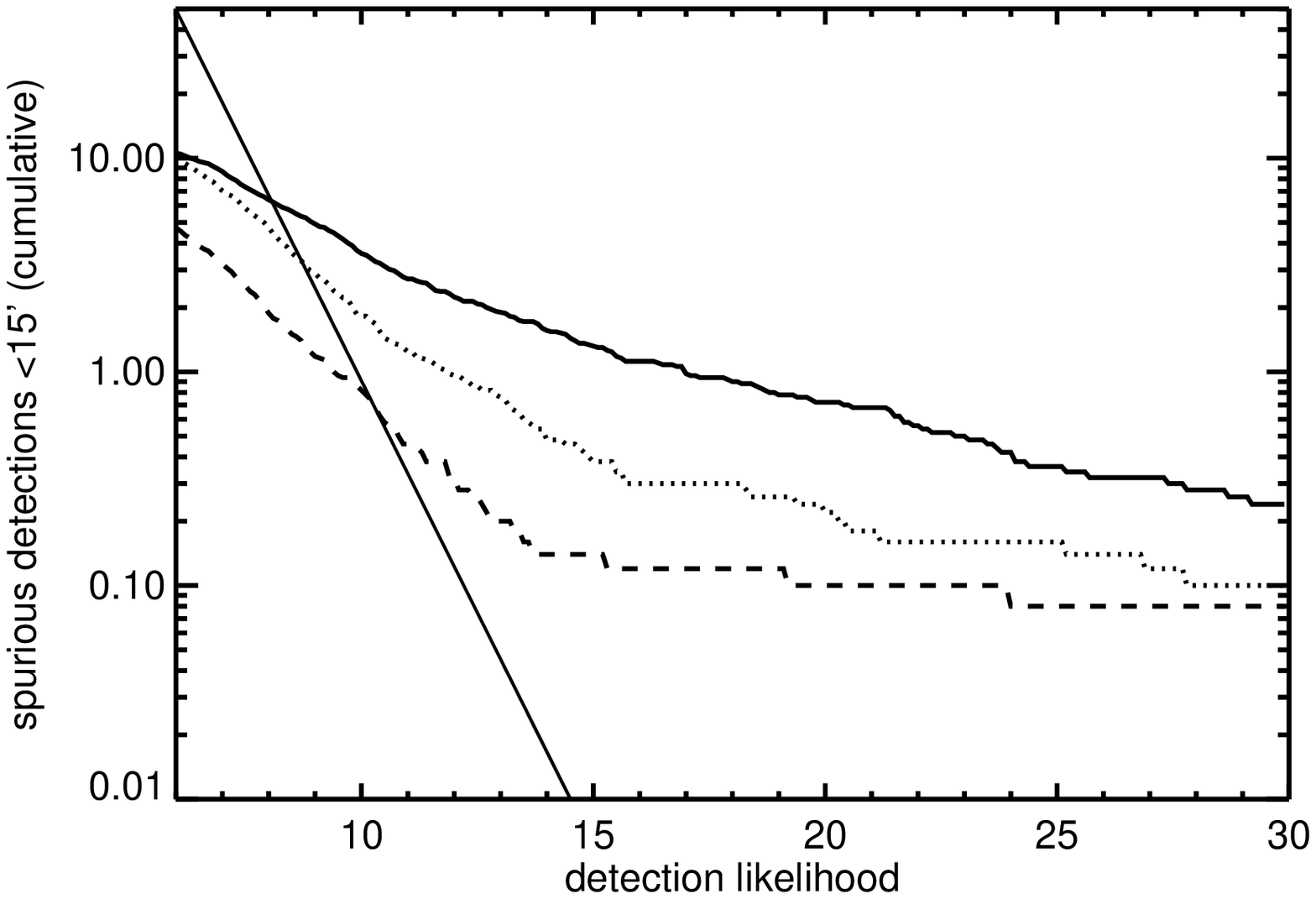}}
\resizebox{\hsize}{!}{\includegraphics{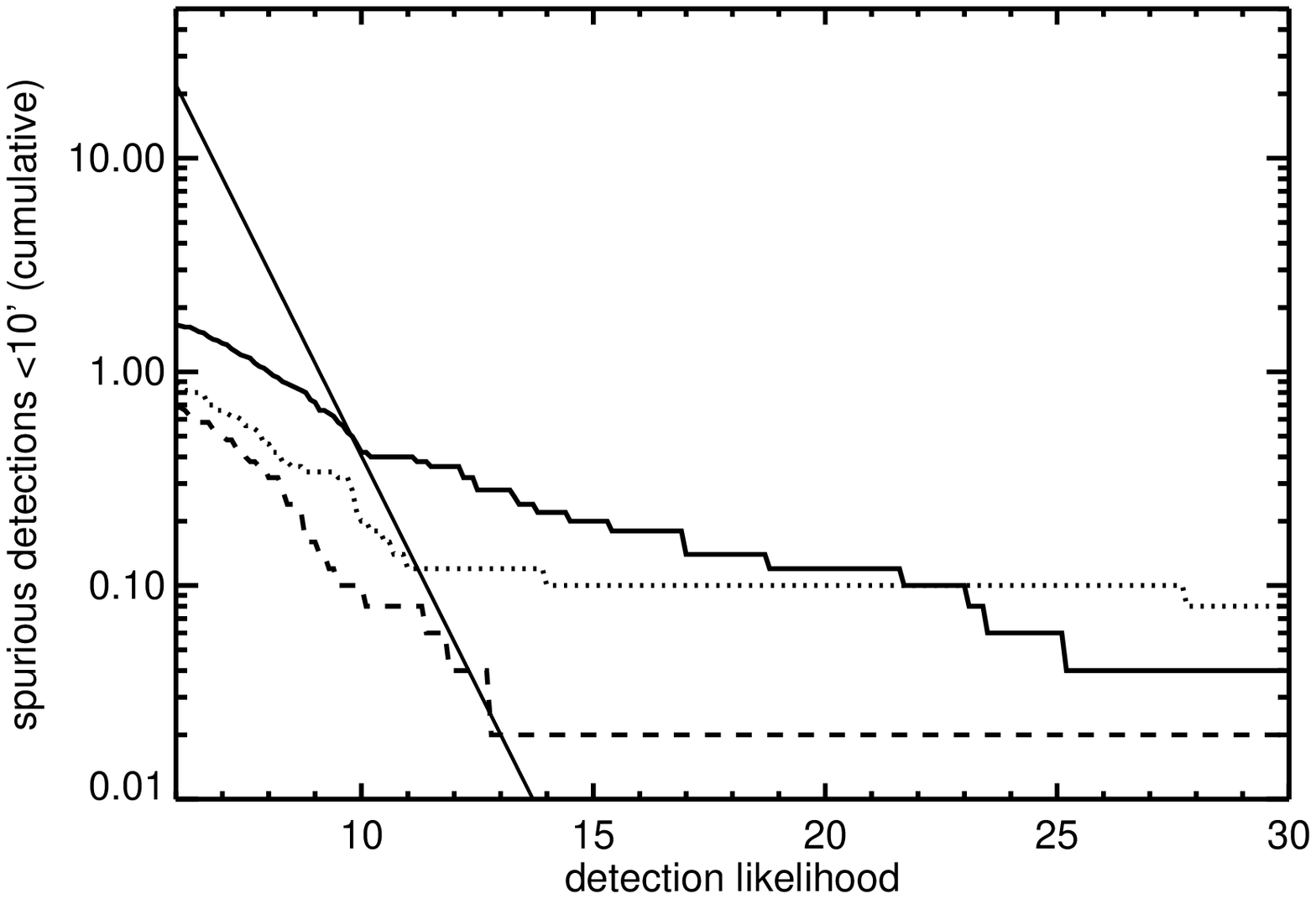}}
\caption{Number of spurious detections from simulations as a function of
 likelihood. {\em Upper~panel:}  spurious detections within 15$^\prime$ around
the field centre. {\em Lower~panel:} the same as above but in the area of 
optical follow-up observations within 10$^\prime$  off the field centre. 
The soft, hard, and very hard energy 
bands are represented by solid, dashed, and dotted lines, respectively. 
The straight lines correspond to the expected number of spurious detections 
based on the detection likelihood as defined in section 4.1, assuming 
20000 (upper panel) and 8900 (lower panel) independent detection cells in the field.}
\label{spurDet}
\end{figure}

\begin{figure}
\resizebox{\hsize}{!}{\includegraphics{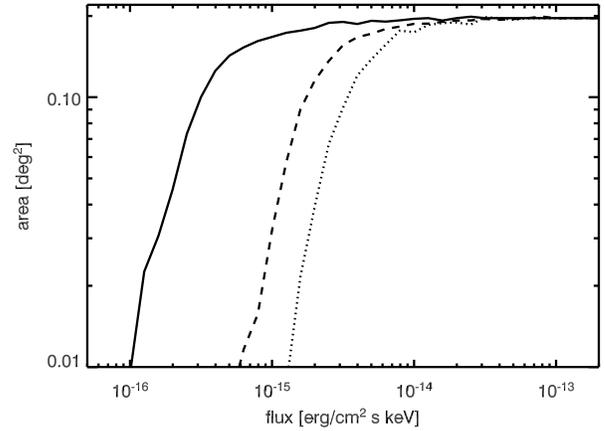}}
\caption{Effective survey area 
as a function of detection sensitivity. The soft, hard, and very hard energy 
bands are represented by solid, dashed, and dotted lines, respectively. }
\label{area}
\end{figure}

The behaviour of the detection procedure was verified by Monte Carlo 
simulations.  We created 50 simulated fields in the soft, hard, and very hard energy band, 
each with the same sky exposure 
distribution and background level, with random source populations 
following a standard logN-logS distribution and the same spectral model as 
assumed for computing the exposure maps.\footnote{ Changing the spectral index by $\Delta \Gamma$=0.3 would 
result in only a few  percent difference in the number of the simulated  sources at the 
limiting flux.}
In order to properly 
model the source  confusion, sources were simulated  down to a soft flux of 
3$\times$10$^{-19}$ erg cm$^{-2}$ s$^{-1}$.
The point spread function
of the simulated sources was constructed from ray-tracing derived 
 point spread function maps available
in the XMM-{\em Newton} calibration database and by assuming the same pointing 
pattern as in the observations. The simulated point spread  function
thus closely matches the observed one at each image location.
We subjected the simulated fields to the same source detection 
procedure as used for analyzing the observations. 
The 50 output catalogues were matched independently with 
the respective input catalogues. Sources with no corresponding simulated input were classified as spurious.
We obtained the fraction of spurious sources by summing the number of spurious sources 
in each output catalogue and by normalizing to the total number of detected sources.
Fig.~\ref{spurDet}
shows the number of spurious detections for sources within
a radius of 15$^\prime$ and 10$^\prime$ around the field centre, respectively. The continuous, 
dashed, and dotted lines refer to detections in the soft, hard, and 
very hard bands; the straight line corresponds to the expected number 
of spurious detections based on the detection likelihood as defined in
section 4.1 and assuming
20000 trials in the full field and 8900 trials within a radius of 10$^\prime$ which
roughly corresponds to the number of independent detection cells 
assuming a  point spread function size of 13$^{\prime\prime}$  half energy width. 
The deviation of the results of the Monte Carlo simulations
from the straight line can be understood in terms of the dual 
detection thresholds of our multi-step detection procedure as well 
as due to the simultaneous Maximum Likelihood fitting of source positions
and fluxes, both of which result in a reduction of the effective number
of independent trials.
It can be seen that most spurious detections occur outside a radius of
10$^\prime$ where the mean point spread function begins to broaden and assumes an
elongated shape.  
Based on the results of the Monte Carlo analysis, we therefore decided to restrict
our source catalogue to sources with detection likelihoods above  10,
corresponding to  up to 4 expected spurious detections within a 
radius of 15$^\prime$ and less than one within
the inner 10$^\prime$, the area used for optical follow-up work.
We also derived, for each energy band, the sky coverage as a 
function of the detection sensitivity, from the fraction of 
simulated input sources detected in each flux interval (Fig.~\ref{area}).
              
\subsection{Number counts}

We computed the logN-logS relations in the soft,
hard, and very hard energy bands using the sky coverage as determined from our Monte Carlo simulations
(see Fig.~\ref{area}) and by estimating the fluxes in the soft, hard, and very hard bands from 
the measured fluxes, assuming a power-law spectral index of $\Gamma=1.4$. 
This spectral index was chosen for estimating the logN-logS fluxes
to directly compare our source number counts to the results
published in the literature. Most the deep X-ray surveys use this spectral index to 
model the spectra of the sources 
to match the spectral shape of the  X-ray background. Changing the spectral index by $\Delta\Gamma\sim$0.3 results in
a flux change of $<$3\% in the soft band and by $<$8\% in the hard and very hard energy  bands. 
Since the sky coverage was derived by realistic Monte Carlo simulations, distortions introduced by the presence of spurious sources in the catalogue are taken into account. In the same way we correct for the Eddington Bias.

\begin{figure*}[!t]
\begin{center}

{\includegraphics[scale=0.4]{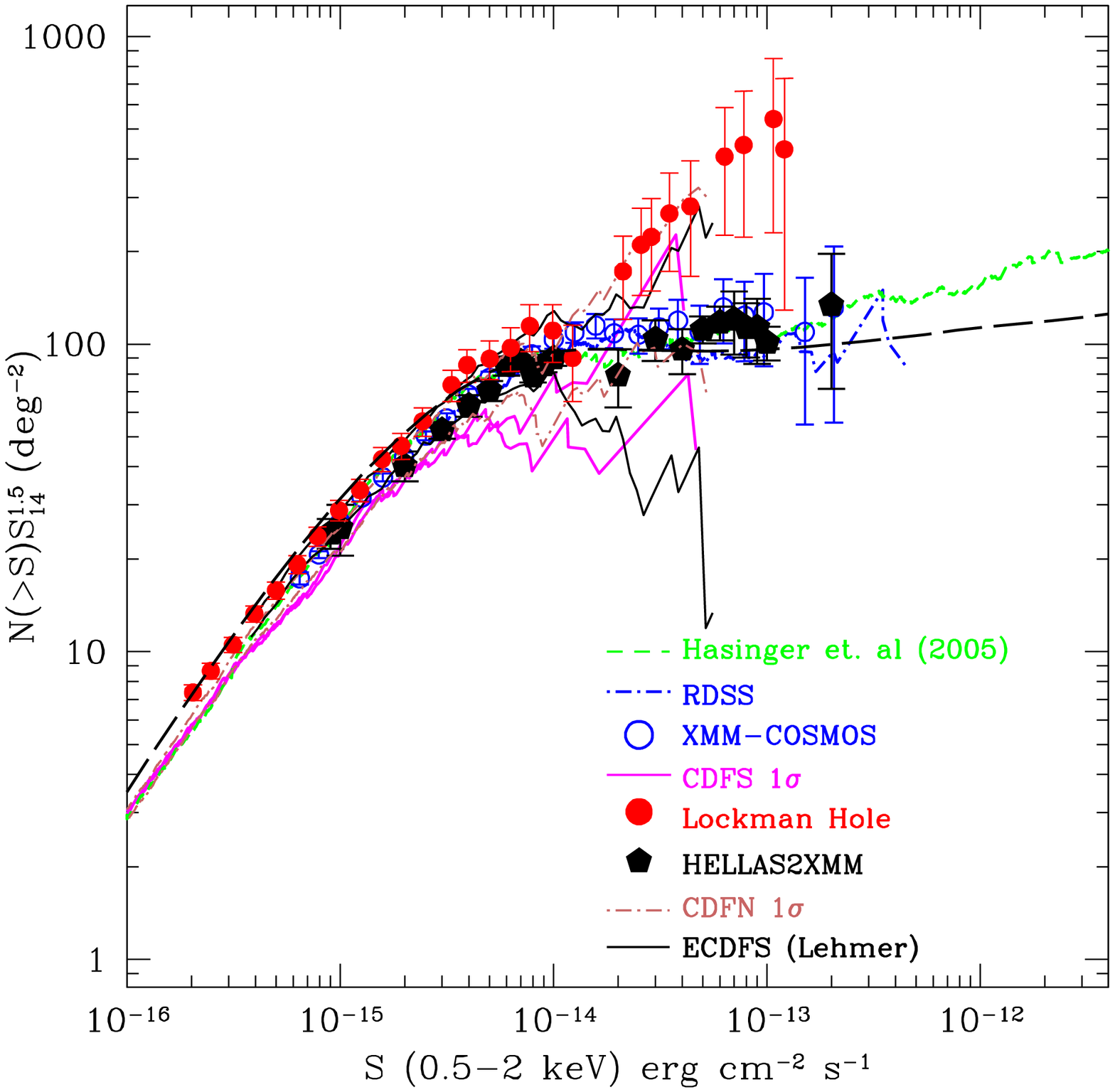}}
{\includegraphics[scale=0.4]{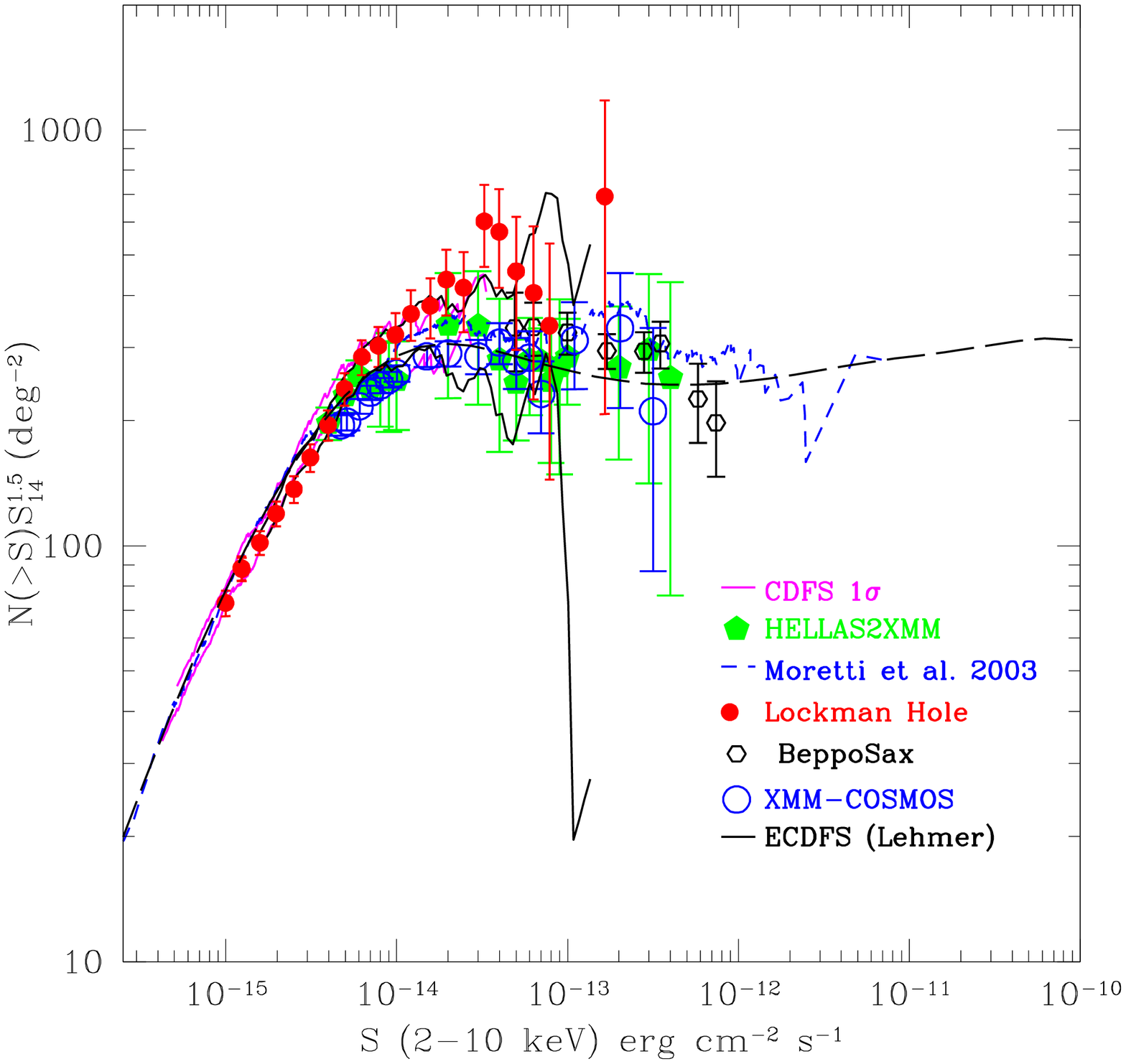}}
{\includegraphics[scale=0.4]{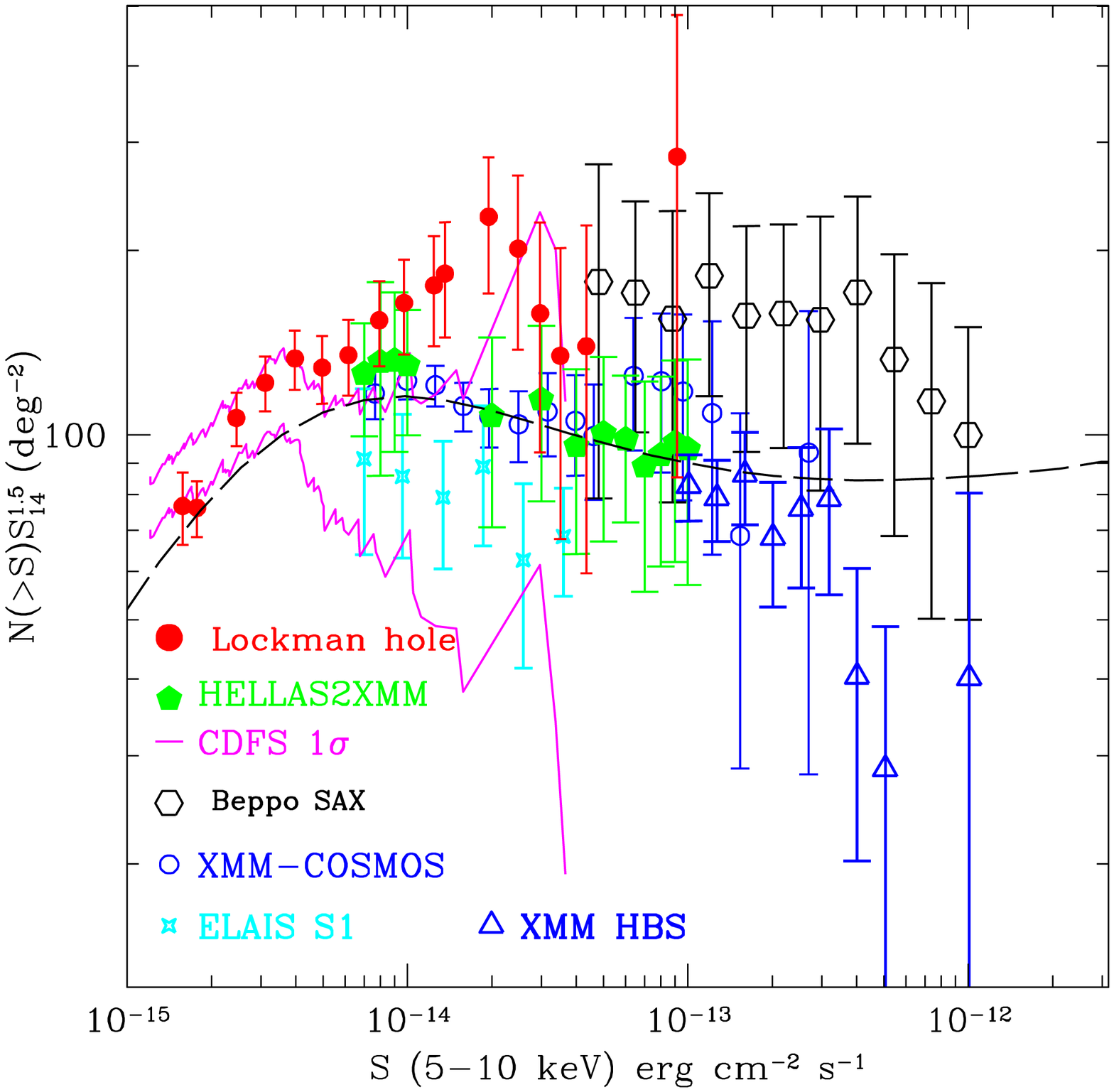}}
{\includegraphics[scale=0.4]{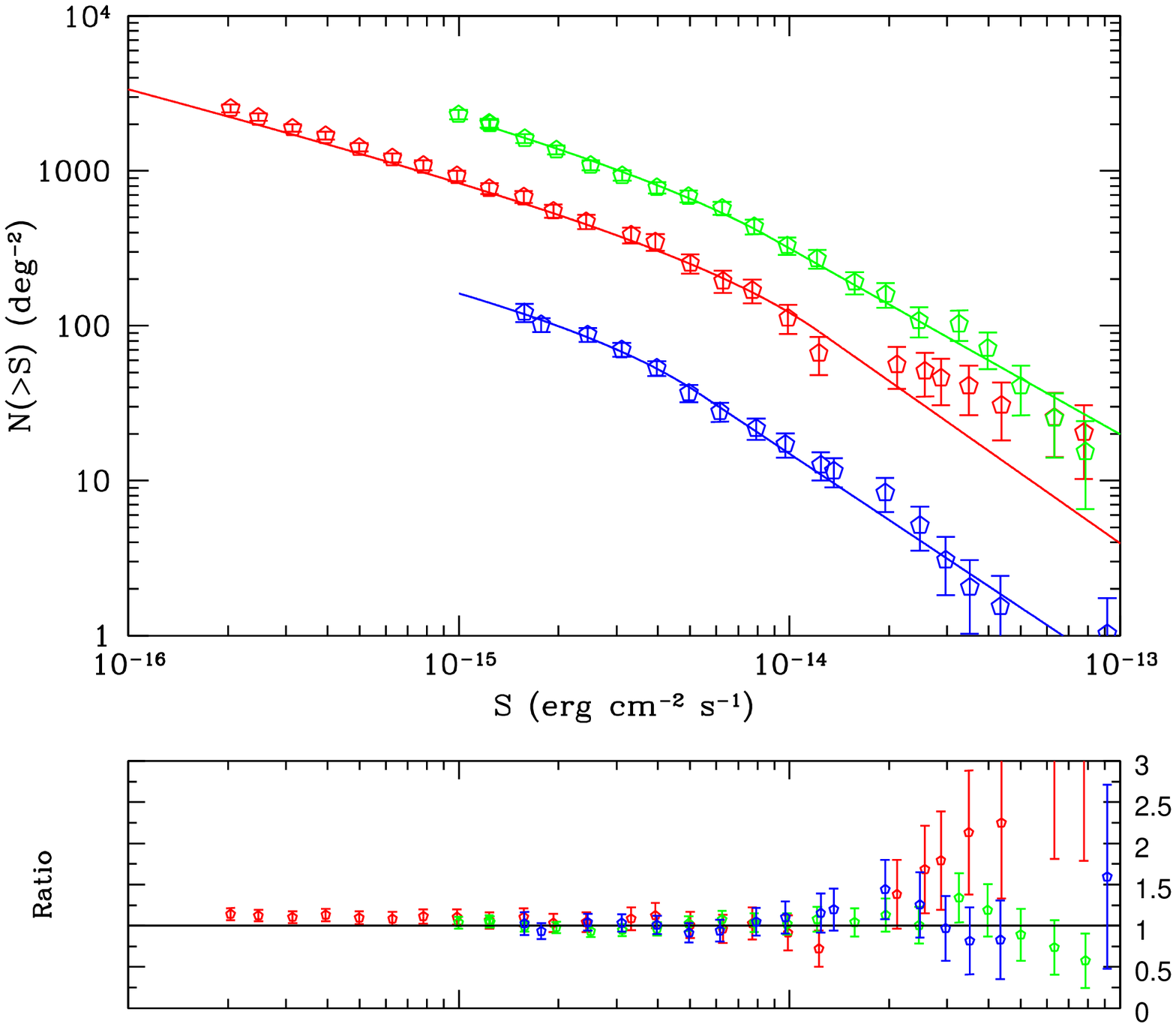}}
\caption{\label{logNlogS}logN-logS relations multiplied by S$^{1.5}$ in the  
0.5--2.0~keV ({\em top~left}),
2.0--4.5~keV ({\em top~right}) and 4.5--10.0~keV ({\em bottom~left}) energy bands.
The data are compared to those obtained in the XMM-COSMOS survey (Cappelluti et al. \cite{nico}),
the HELLAS2XMM survey (Baldi et al. \cite{baldi}),
the compilations of X-ray surveys produced by 
Moretti et al. (\cite{more}) and Hasinger et al. (\cite{hasinger05}),
 the Beppo-Sax survey (Fiore et al. \cite{fiore}, Giommi et al. \cite{giommi}),
the {\em Chandra} deep surveys CDFN and CDFS (Rosati et al. \cite{rosati}, Bauer et al. \cite{bauer}),
the RDSS (Hasinger et al. \cite{hasinger93}),
the extended CDFS (Lehmer et al. \cite{lehmer}), the ELAIS-S1 survey (Puccetti et al. \cite{pucc}) and 
the XMM-HBS survey (Della Ceca et al. \cite{rdc}). 
The black dashed lines represents the prediction of the X-ray background model of Gilli et al. (\cite{gilli}).
\label{fits}{\em Bottom~left}:  in the {\em upper~panel} the X-ray  logN-logS in the Lockman Hole field 
({\em pentagons}), and the broken power-law fits ({\em solid~lines}) in the soft, hard, and very hard 
energy band, in {\em red}, {\em green}, and {\em blue}, respectively. For clarity the very hard band relation 
is divided by ten. In the {\em lower~panel} data/model ratio, colour code as in  the {\em upper~panel}.}
\end{center}
\end{figure*}

Objects down to a detection likelihood of 6
in the respective energy band were considered for the determination of the 
number counts, corresponding to 5 to 10 expected spurious detections
based on results from our Monte Carlo simulations (Fig.~\ref{spurDet}). 
If $\Omega_{i}$ is the sky coverage at a flux limit {\em S}$_{i}$, the
cumulative number counts are obtained via: 
\begin{equation}
N(>S_{i})=\sum_{i=1}^{N_{sou}} \frac{1}{\Omega_{i}}.
\end{equation}
The logN-logS relations are plotted in Fig.~\ref{logNlogS}, 
multiplied by S$^{1.5}$ in order to enhance deviations
from the Euclidean behaviour. The logN-logS are compared with
a sample of XMM-{\em Newton}, {\em Chandra}, ROSAT, and Beppo-Sax surveys.

We  performed a  Maximum Likelihood fit to the unbinned differential logN-logS.
As commonly used in the literature, we assumed a broken power-law of the form
 \begin{equation}
 n(S)=\frac{dN}{ds}= \left\{\begin{array}{ll}
A\,S^{-\alpha_{1}} & S>S_{b} \\
B\,S^{-\alpha_{2}} & S \leq S_{b},\\
\end{array}
\right.
\end{equation} 
where $B=A\,S^{\alpha_{2}-\alpha_{1}} $. $\alpha_{2}$ and $\alpha_{1}$ are the faint and
bright slopes, respectively.
 $S$ and the cut-off flux {\em S}$_b$ are  expressed in units of 10$^{-14}$ erg cm$^{-2}$ s$^{-1}$ and 
A is the differential normalization in units of deg$^{-2}$ at 10$^{-14}$ erg cm$^{-2}$ s$^{-1}$.
 The likelihood function $\mathcal{L}$, has been  defined as in Murdoch, Crawford \& Jauncey (\cite{mur}). The best-fit parameters $\alpha_{1_{best}}$, $\alpha_{2_{best}}$ and {\em S}$_{b_{best}}$ satisfy the condition 
$-2\mathcal{L}(\alpha_{1_{best}},\alpha_{2_{best}},S_{b_{best}})$=min, the confidence intervals 
are defined to be $\Delta \mathcal{L}=\mathcal{L}(\alpha_{1},\alpha_{2},S_{b})-\mathcal{L}_{best}$
where $\mathcal{L}_{best}$ is the minimum likelihood.
Since $\mathcal{L}$ is distributed like $\chi^{2}$,  for 3 interesting parameters, 
the 1$\sigma$ confidence interval is $\Delta \mathcal{L}$=3.53. 
In the soft band because of the low statistics at the bright end of the logN-logS,
we fixed $\alpha_{1}$ to 2.50;  the best-fit value of the free parameters are $\alpha_{2}$=1.55$\pm{0.05}$, 
{\em S}$_{b}$=1.23$\pm{0.40}$ $\times$ 10$^{-14}$~erg~cm$^{-2}$~s$^{-1}$ and A=187 deg$^{-2}$.
In the hard band the best-fit parameters are $\alpha_{1}$=2.20$\pm{0.16}$,  $\alpha_{2}$=1.55$\pm{0.12}$, 
{\em S}$_{b}$=0.88$\pm{0.18}$ $\times$ 10$^{-14}$~erg~cm$^{-2}$~s$^{-1}$ and A=379 deg$^{-2}$.
In the very hard band we obtained $\alpha_{1}$=2.42$\pm{0.18}$,  
$\alpha_{2}$=1.41$\pm{0.34}$, {\em S}$_{b}$=0.53$\pm{0.10}$ $\times$ 10$^{-14}$~erg~cm$^{-2}$~s$^{-1}$ and A=212 deg$^{-2}$.
The results of the fit are summarized in table~\ref{llFits}. In order
to visualize the results of the fits, in Fig. \ref{fits} we plotted the logN-logS  relations 
and their best-fit broken power-laws and the data/model ratios in the lower panel. As can be noticed, the faint end part is well represented by our fits in all the energy bands.
 In the soft band, where  the parameter $\alpha_{1}$ was frozen, the fit underestimates the source number counts. This effect is caused by the well known overdensity of bright X-ray sources in the Lockman Hole resulting in a flattening of the relation at high fluxes. This effect is discussed in detail in Section 5.   In the other energy bands our fits reproduce well the behaviour of the logN-logS relations also at the bright end side, though the bright end slopes remain slightly flatter that the canonical $\alpha_{1}$=2.5.

\begin{table}
\begin{center}
\caption[]{ Best-fit parameters for logN-logS relation of eq. 2}
\label{llFits}
\begin{tabular}{lllll}
\hline
energy band & A$^a$& $\alpha_{1}^b$ & $S_{b}^c$ & $\alpha_{2}^{d}$  \\ \hline
0.5--2.0 keV & 187 &  2.50$^{\star}$& 1.23$\pm{0.40}$ & 1.55$\pm{0.05}$   \\
2.0--10.0 keV & 379 & 2.20$\pm{0.16}$& 0.88$\pm{0.18}$ &  1.55$\pm{0.12}$\\
5.0--10.0 keV&  212 &  2.42$\pm{0.18}$ & 0.53$\pm{0.10}$ &   1.41$\pm{0.34}$      \\ \hline 
\end{tabular}
\begin{list}{}{}
\item[$^a$] Normalization  in units of deg$^{-2}$ at 
            10$^{-14}$ erg cm$^{-2}$ s$^{-1}$
\item[$^b$] Bright end slope
\item[$^c$]  Cut-off flux in units of 10$^{-14}$ erg cm$^{-2}$ s$^{-1}$
\item[$^d$] Faint end slope 
\item[$^\star$] fixed

\end{list}
\end{center}
\end{table}
  
\subsection{Source catalogue}


\begin{figure}
{\includegraphics[width=12cm]{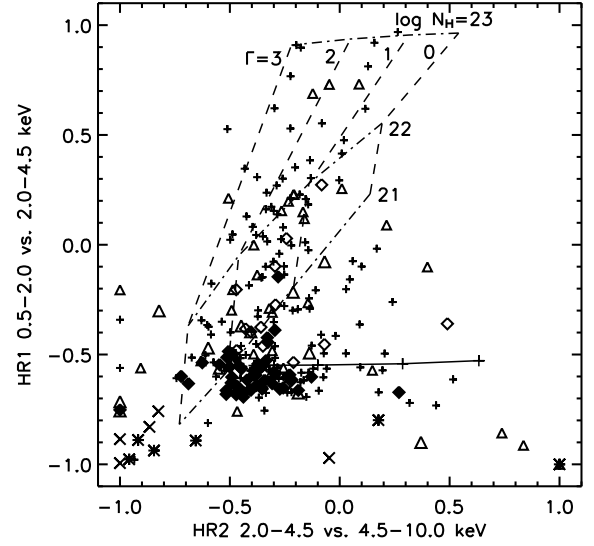}}
\caption{\label{colCol}X-ray colour-colour plot.
Hardness ratios of objects with  identified counterparts 
(filled diamonds: Type-I AGN, open diamonds: Type-II AGN, Open triangles: normal galaxies, 
crosses: clusters and  groups, stars: stars) and for objects (+) 
where both hardness ratios have 1$\sigma$ errors $<$ 0.25. 
Error bars were omitted for clarity; typical 1$\sigma$ errors are in the range 0.05--0.25
(see table~\ref{souCat}). Grid lines: location of objects with different absorbed power-law spectra.
 Dashed lines: location of objects with spectral index $\Gamma$ = 0 to 3;  Dot-dashed lines: photoelectric 
absorption levels from log(N$_{\rm H}$)=21 to log(N$_{\rm H}$)=23. Solid line:  reflection+leaky absorber model, see
text for details.}
\end{figure}

In order to obtain a source catalogue which is largely free of spurious detections
we limited the catalogue to objects with a full band  detection
likelihood $>$ 10, corresponding to  up to 4 spurious detections based on our
Monte Carlo simulations,  which is $\sim$1 \% of the detected objects.
Table~\ref{souCat}   
lists the detection likelihoods,  fluxes, and flux errors (1$\sigma$) in the total band, as well
as in each of  the three individual bands. The fluxes were calculated from the combined
EPIC pn and MOS count rates determined by Maximum Likelihood fitting of the
point spread function and assuming a power-law spectral index of $\Gamma=2.0$.
While source detection was performed in the
three non-overlapping energy bands 0.5--2~keV, 2--4.5~keV, and 4.5--10~keV, 
source fluxes are quoted in the soft, hard, and very hard band (see Section  3) to better compare our data with published results. 
The total band fluxes were calculated by summing the soft and hard band fluxes.
For a fraction of the sources
significant detections were only achieved in a subset of the energy bands. 
Individual band flux values with corresponding detection likelihood values of less 
than 10 should therefore  not be regarded as significant detections.
 The flux errors are the statistical uncertainties
estimated from the Maximum Likelihood and do not include uncertainties introduced by the choice of the spectral
model to estimate the flux. In Section 4.3 we  estimated the variation of the flux  by changing the spectral
index by $\Delta \Gamma$=0.3, this is $\sim$3\% in the soft band and $\sim$ 8\% in the hard and very hard bands. 
In total 340, 266, and 98 
sources were detected in the soft, hard, and very hard band, respectively  corresponding to a 
total of 409  unique sources. 
There are 117 sources which were only detected in the soft band and
37 sources only detected in the hard  band. All of the 98 sources
detected in the very hard band were also seen in the hard band. Six of the
objects detected in the very hard band were sufficiently absorbed 
such that they
were not visible in the soft band.
We have assigned unique numbers\footnote{The numbering scheme was chosen for internal, technical reasons; it does not reflect any of the tabulated source properties.} to each of the detected sources. 
 In the background subtracted full band image of Fig.~\ref{LHimg} 
each detected source is marked by its 1$\sigma$ positional error 
circle and source number.
The ROSAT Lockman Hole catalogue (Lehmann et al. \cite{lehmann}) 
was searched for counterparts in the XMM-{\em Newton} catalogue. 
ROSAT sources  falling within 6$^{\prime\prime}$ from the XMM-{\em Newton} centroid 
were identified as previously observed and their ROSAT source number is given.
For most of these objects the 
ROSAT and XMM-{\em Newton} soft band fluxes were found to be broadly in 
agreement within statistical errors and considering the somewhat different 
spectral response of the instruments, with an indication for variability 
of up to a factor of $\sim2.5$ in a subset of the objects.

We list two hardness ratios for each source, defined as

\[ HR_1 = \frac{B_2 - B_1}{B_2 + B_1} {\rm~~~~~~and~~~~~~} HR_2 = \frac{B_3 - B_2}{B_3 + B_2} \]

where $B_1$, $B_2$, and $B_3$ refer to the vignetting corrected count rates,
summed over all three XMM-{\em Newton} cameras,
in the soft, hard, and very hard energy bands, respectively.
 Statistical 1$\sigma$ hardness ratio errors were calculated from the 
count rate errors by error propagation.
In Fig.~\ref{colCol} the hardness ratios are displayed in the form of an X-ray
colour-colour plot. The plot contains 259 objects 125 of which
have known spectroscopic classifications, both from optical follow-up
of previous ROSAT observations (69 objects; Lehmann et al. \cite{lehmann}) 
and from the ongoing optical follow-up work of the XMM-Newton observations (Szokoly et al., in preparation).
 The remainder of the objects are as yet unidentified
XMM-Newton detections where the statistical 1$\sigma$ error of both
hardness ratios  is less than 0.25 (134 objects). The objects with large hardness
ratio errors  are omitted for display  purposes only; except for their larger scatter, 
we did  not see any obvious deviation
of the distribution of their X-ray colours from the displayed objects.
Filled and  open diamonds refer 
to Type-I and Type-II AGN, respectively. Galaxies are shown
as open triangles, clusters and groups are displayed as crosses (X), and stars 
are shown as stars. The small  plus signs
correspond to as yet unidentified XMM-{\em Newton} objects. 
 The grid lines refers to spectral models that were folded through 
the instrument response of the pn+MOS1+MOS2 detectors, using XSPEC. Each grid line corresponds
to a simple power-law spectrum with photon indices $\Gamma=0,1,2,3$ (from right to left) 
and intrinsic absorption (in the observer frame) of 
N$_{\rm H}=10^{21}$, $10^{22}$ and $10^{23}$ cm$^{-2}$ (from bottom to top). 
While most Type-I AGN tightly cluster in one location consistent with a standard
AGN-type power-law spectrum with very little absorption,  Type-II AGN fill 
most of the HR$_1$-range, corresponding to observer frame absorption up to 
$10^{23}$ cm$^{-2}$.

A small number of predominately Type-II AGN is found to have HR$_1$ hardness 
ratios typical of Type-I AGN but relatively hard corresponding HR$_2$ values, 
inconsistent with a single absorbed power-law. 
Our hypothesis is that these sources are Compton thick (N$_{\rm H}>\sim$ 1.5$\times$10$^{24}$ cm$^{-2}$) 
objects, whose spectrum is dominated by a Compton-reflection continuum 
from a cold medium which is usually assumed to be produced by the inner 
side of the putative obscured torus plus a soft power-law which is made 
by the photons "leaking" through the absorber. We used the 
XSPEC model "pexrav" (Magdziarz \& Zdziarski 1995) to reproduce such a 
spectrum and  to compute the expected hardness ratios.
The track in Fig. \ref{colCol}  has been obtained for  a leaking flux going from 1 to 
30\% of the total flux observed, from right to left. The increase of the 
leaking flux dilutes more and more the signature of the reflection 	
component, modifying therefore the X-ray  colours. The fraction of the 
flux due to reflection and leaking flux is described by the parameter 
"rel\_refl"  of the pexrav model.	
These objects may thus be representatives 
of a class of very highly absorbed AGN which due to the soft flux leaked through a
partially covered absorber are detectable by XMM-{\em Newton}.

A source table giving the full details of the detection results is available in electronic form.  

A separate source detection run was performed where sources were tested for significant 
extent beyond the size of the point spread function, assuming a King profile  
(see details in section  4.1). 13 objects were found with significant extent
(likelihood of extent $>$ 15, corresponding to  5$ \sigma$) with best-fit King profile 
extent values in the range from 1.5$^{\prime\prime}$ to 20$^{\prime\prime}$. 
Three objects where the best-fit extent reached
the maximum permitted value of 20$^{\prime\prime}$ were discarded as experience with the
detection software shows that such objects tend to be spurious.
In addition, two objects with extent sizes below 1$^{\prime\prime}$ were assumed to be artifacts of
inaccuracies in the modeling of the point spread function and were discarded 
from the list of extended objects. Table~\ref{souCatExt} lists the source positions,
the likelihood of detection and likelihood of extent, and the best-fit King profile
extent in arcsec, as well as source fluxes and hardness ratios. Where the detected extended
objects overlapped with detections from our list of point sources, the source numbers
from the point source list are given. Only one extended object was not also detected in
our point source search. 
Of the seven ROSAT objects within the area covered by the current analysis,
classified as either groups or clusters of galaxies, five were
detected as extended by XMM-{\em Newton}. The remaining two objects appear
pointlike on the XMM-Newton image and were only detected as point sources. 
One object classified as Type-II AGN in Lehmann et al. (\cite{lehmann}) 
was found to be marginally extended.
\section{Discussion}
  It is known from the literature that the shape of the extragalactic 
X-ray logN-logS is well constrained (see  e.g. Kim et al. \cite{kim} and references
therein). The normalizations of such  relations show a wide field to field 
variation  (see e.g. Cappelluti et. al \cite{nico05,nico},  Yang et al. \cite{yang} and 
references therein). These fluctuations have been explained by superposition
of Poissonian noise and fluctuations introduced by the clustering of the sources. 
The relative importance of these two components strongly depends on the depth and the 
sky coverage of the survey. In  pencil beam surveys there is a higher chance
of detecting strong variations, introduced at the bright end by the low  count statistics
and at the faint end by chance sampling of local large scale structures. 
In section 4.3 we presented the logN-logS relation for the X-ray sources 
in the XMM-{\em Newton} Lockman Hole survey  in the context of previous results (see Fig.~\ref{logNlogS}).
In order to compare our results with previously published work we performed  a comparison
of the best-fit parameters together with a visual inspection of the plots.

\begin{itemize}
\item At fluxes lower than 10$^{-14}$ erg cm$^{-2}$ s$^{-1}$,
the soft band  logN-logS relation shows an excellent agreement with previous surveys. 
Also, the best-fit parameters in this flux region are in agreement with results 
presented in previous surveys and summarized in Kim et al. (\cite{kim}). 
As an example, the value of $\alpha_{2}$=1.55$\pm0.05$ measured in this work is 
consistent within 1$\sigma$ with  the results of  Hasinger et al. (\cite{hasinger05}) who obtained
$\alpha_{2}$=1.55$\pm0.04$, and with those of Moretti et al. (\cite{more}) who measured 
$\alpha_{2}$=1.57$\pm0.10$. Both of the above cited works have been performed 
by merging the catalogues and the sky coverages of most of the {\em Chandra}, XMM-{\em Newton} and ROSAT 
surveys. 
 Interestingly enough, at fluxes lower than {\em S}$_{b}$, our logN-logS is in good agreement also with the prediction of Gilli et al. (\cite{gilli}) based on  X-ray background population synthesis. 
At the bright end of the distribution we detected an excess of sources. 
This  is  known to be a field selection effect of the Lockman Hole introduced to
improve the ROSAT attitude precision (see e.g. Hasinger et al. \cite{hasinger98}).
Such an overdensity is however not statistically significant, and following Cappelluti et al. (\cite{nico}),
this can be  explained by Poisson fluctuations in the low source surface density regime.
Without considering the bright end excess it is worth noting that the normalization
measured in this work is in excellent agreement with those measured in the XMM-COSMOS field
which, being obtained on an area of $\sim$2 deg$^{2}$, can be considered as not affected by cosmic variance at a level $>5\%$.  
 
\item In the hard band we were able to constrain all the parameters of the logN-logS
      and, even in this band, our results agree with most of the previous survey
      results listed in table 4 of Kim et al. (\cite{kim}). 
      Also in this band the best-fit parameters are in excellent agreement with
      those of the XMM-COSMOS survey with smaller errors on  $\alpha_{2}$ due to
      a better sampling of the flux interval below the knee of the relation. 
      A direct comparison with the compilation of Moretti et al. (\cite{more})
      who obtained $\alpha_{2}$=1.44$\pm{0.13}$ also shows a 1$\sigma$ consistency.
      The bright end slope, $\alpha_{1}$=2.2$\pm{0.12}$ measured in this work is somewhat 
      flatter than the expected Euclidean 2.5. This can be explained with the overdensity 
      of bright sources at the bright end as observed in the soft band. 	
\item In the 5--10~keV band the flux range of the XMM-Lockman Hole survey allows us to 
      sample with very good statistics the regions above and below {\em S}$_{b}$. 
      The position of the break is in good agreement with the predictions
      of Gilli et al. (\cite{gilli}). We were able, for the 
      first time, to constrain $\alpha_{2}$ in this band but with  quite 
      large uncertainty.  The normalization shows a $\sim$40\% higher
      value than in the XMM-COSMOS survey. This band is in fact more sensitive
      to the bright source overdensity. 
      Since all the bright sources in the soft band source excess are detected in the very 
      hard band,  the fraction  of bright sources which reside in the source excess  
      is higher in this band than  in the soft and hard energy bands. A consequence of this 
      peculiarity is therefore an increase of the normalization of the logN-logS along
      the totality of the relation. At the very faint end in fact the overdensity gets weaker. 
      The points of the logN-logS are in agreement with the measurement of the CDFS, 
      HELLAS2XMM and partly with the  X-ray background model of   Gilli et al. (2007). 
      Moreover, with the only exception of the CDFS, whose
      points are in agreement with ours below the knee,
      all the surveys derive their fluxes using $\Gamma$=1.7 as model spectral index. 
      This has the effect of reducing the normalization, in this band,  by $\sim$15\%.
 \end{itemize}     
        The contribution of the Lockman Hole X-ray sources to the  X-ray background has been 
         discussed in detail by Worlsey et al. (\cite{worlsey05}) resulting in important conclusions  
  	on the nature of the missing sources of the X-ray background. 
	The distribution of the different types of sources contributing to the 
         X-ray background can, in the first instance,
	be determined by carefully examinating the colour-colour diagram. Type-I AGN are 
	clustered in a well defined region, corresponding to low intrinsic absorption 
	and  photon indices in the range 1.7--2 as determined by Mainieri et al. (\cite{mainieri}).
	The bulk of the distribution of  Type-I AGN, seems to be right-shifted from the no absorption
	locus, suggesting the presence of  additional spectral components. 	
	On the other hand, Type-II AGN show higher hardness ratios,
	corresponding to  higher intrinsic absorption. 	  		
	As pointed out  by Hasinger et al. (\cite{hasinger06}) in the COSMOS field, 
	there is evidence of candidate Type-II AGN with quite hard colours in the hard bands  (HR$_2$),
	and soft  band colours  (HR$_1$) consistent with unabsorbed sources. The solid horizontal line in Fig. \ref{colCol} shows however that their colours are  similar to those of heavily absorbed sources at low redshift, with a small
	fraction of unabsorbed flux leaking out. 
	
\begin{figure}[!t]
\center
\resizebox{\hsize}{!}{\includegraphics{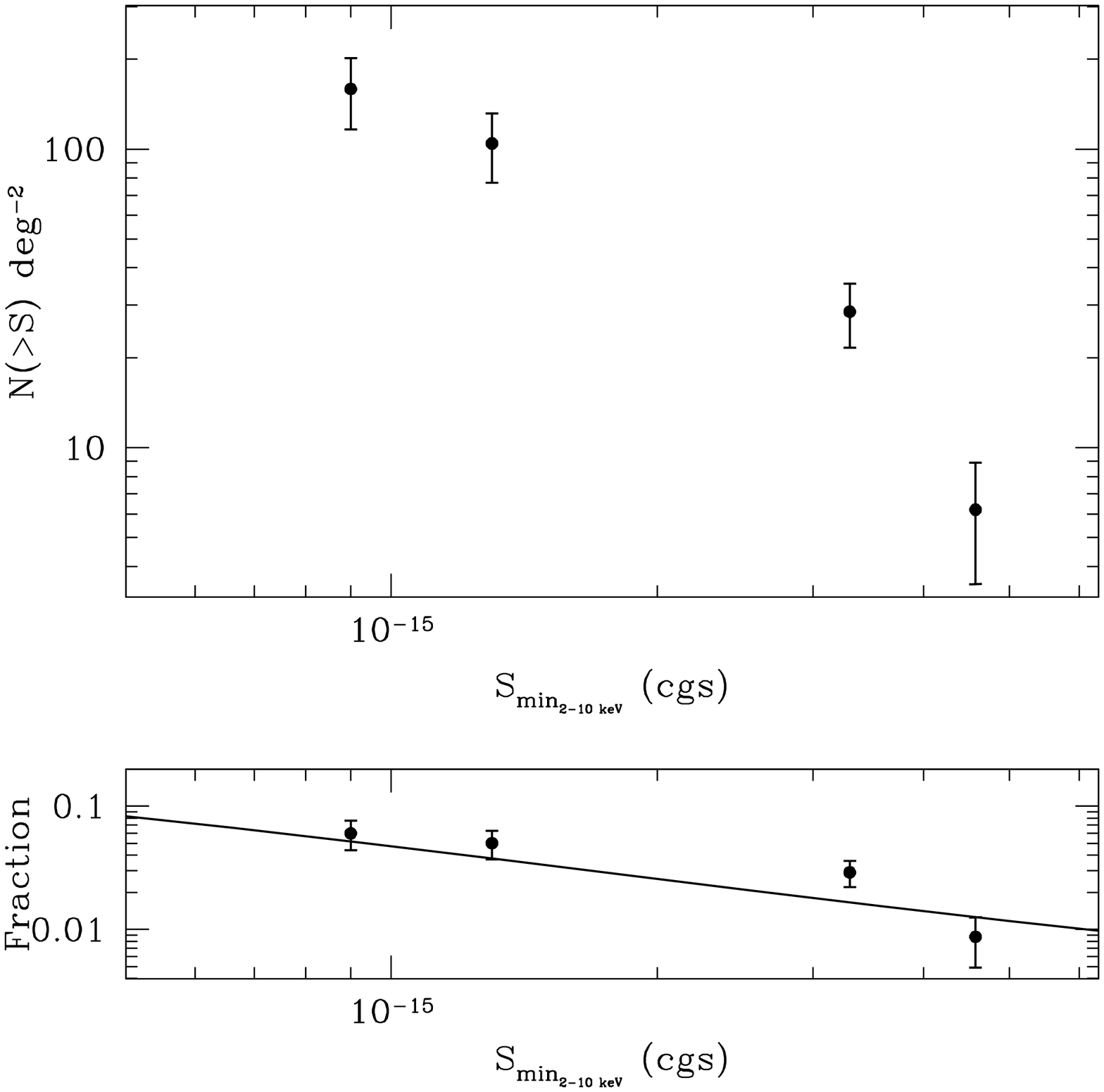}}
\caption{\label{fig:cthick}{\em Upper~panel:} The logN-logS of Compton thick source candidates as derived from the observed
fraction rescaled to the expected number density at  the flux of faintest source of the survey. 
From left to right: the Lockmann Hole (this paper), the CDFS (Tozzi et al. 2006), 
the XMM-COSMOS (Hasinger et al. 2007) and the 
{\em Chandra}-SWIRE survey (Polletta et al. 2006). 
{\em Lower~panel:} The fraction of  Compton thick source  candidates in the Lockmann Hole, 
the CDFS, the XMM-COSMOS and 
{\em Chandra}-SWIRE, respectively from left to right. The solid lines represents the prediction 
of the  X-ray background model
by Gilli et al. (2006). }	 
\end{figure}

	At higher redshift the absorbed 
	continuum moves to softer energy and therefore the source shifts to 
	another locus on the colour-colour plot.
	A single prototype of these objects has been
	detected in the Lockman Hole by Mainieri et al. (2002). The source \#290 ( ROSAT \#901), 
	at redshift $z$=0.204,
	shows an absorbed power-law hard component with  N$_{\rm H}\sim$5$\times$10$^{23}$ cm$^{-2}$
	and an additional soft steep $\Gamma\sim$3 unabsorbed component.  
        A similar  result could be obtained also with the ``pexrav`` XSPEC model.
	Mateos et al. (\cite{mateos}) detected
	sources with high absorption and evidence of soft excess.
	First results from the XMM-COSMOS survey (Hasinger et al. \cite{hasinger06}, 
        Mainieri et al. \cite{mainieri2})
	pointed out that this kind of source could be among the most highly absorbed 
        objects detectable in an X-ray survey.
	 Selecting sources with HR$_1<$-0.1, HR$_2>$0.1 and errors in both hardness ratios less 
         than 0.25, we obtained a sample of 13  candidate Compton thick sources,
	 corresponding to a fraction of $\sim$6\% of Compton thick objects. The 
		faintest among theses sources has  a 2--10~keV flux of $\sim$9$\times$10$^{-16}$ \flux.
		In the {\em Chandra} deep field Tozzi et al. (2006) spectroscopically detected 14/280 	
		candidates at 2--10~keV  fluxes greater that $\sim$1.5$\times$10$^{-15}$ \flux.
             Polletta et al. (2006), detected 5/567 X-ray selected candidates in the {\em Chandra}
	     SWIRE survey at 2--10~keV fluxes greater than $\sim$4.6$\times$10$^{-15}$ \flux.	 	
	     In the XMM-COSMOS survey Hasinger et al. (2007) selected in the same way  as in this
	     	paper 18/600 candidates with fluxes greater than $\sim$3.3$\times$10$^{-15}$ \flux. 
	In the lower panel of Fig. \ref{fig:cthick} we plot the fraction of Compton thick candidates
	as a function of the flux of the deepest candidate sources  in the surveys mentioned
	above.\\
	We note that this determination is in good agreement with   X-ray background model predictions (see e.g. Fig. 16 in Gilli et al. 2007).  Indeed, the fraction of Compton thick sources shows a growth of a factor $\sim$5 
in the 2--10~keV flux interval $\sim$4.6$\times$10$^{-15}$ \flux  to $\sim$9$\times$10$^{-16}$ \flux.  	
	In order to obtain  the first measurement of the surface density of Compton thick sources,
	we  multiplied the source fraction estimated above  in the 2--10~keV  band by the 
cumulative number density predicted by  Gilli et  al. (2007). The resulting logN-logS is shown in the upper 
panel of Fig. \ref{fig:cthick}. 
	Such a logN-logS is well represented by a single power-law  of the form
	\begin{equation}
	N(>S)=2.19\pm{0.88}\times\left(\frac{S}{10^{14}\;{\rm erg\;cm}^{-2}\;{\rm s}^{-1}}\right)^{1.83\pm{0.23}}{\rm deg}^{-2}
	\end{equation} 
	We therefore conclude that the number density of Compton thick sources in  deep XMM-{\em Newton} and 
{\em Chandra}  surveys is rising according to an Euclidean growth. The observed growth of the fraction of Compton thick sources at low fluxes 	is in good agreement  with models. Such a relation, if confirmed at fainter
	fluxes would, in principle, explain the observed shape and the position of the peak of the 
 X-ray background  spectrum.
	 
\section{Conclusion}

We analyzed a set of 18 XMM-{\em Newton} pointings centred on the Lockman Hole which constitute 
the deepest exposure performed by XMM-{\em Newton}, reaching a sensitivity limit of
1.9$\times10^{-16}$ erg cm$^{-2}$ s$^{-1}$,  9$\times 10^{-16}$ erg cm$^{-2}$ s$^{-1}$
and 1.8$\times10^{-15}$ erg cm$^{-2}$ s$^{-1}$, in the 0.5--2.0 keV, 2.0--10.0 keV
and 5.0--10.0~keV energy bands, respectively. 409 sources were
detected within the survey area of 0.20 sq. degrees. 340, 266, and 98 objects
were detected in the soft, hard, and very hard band, respectively. 
A catalogue including the main X-ray characteristics of 
the sources in the Lockman hole survey  is presented.   
The number counts derived from the XMM-{\em Newton} data are in close agreements
 with previous surveys and the most recent synthesis models of the  X-ray background. 
Differences in the  normalization
of our  logN-logS with those of other surveys have been  discussed and explained in terms of low counting
statistics typical of deep pencil beam surveys.
 The high throughput of the XMM-{\em Newton} telescope allowed us 
 to compute and constrain the parameters of the logN-logS
 in the hard and very hard energy 
 bands. This region of the X-ray spectrum is fundamental for understanding 
  absorbed AGNs which are thought to be the most important  contributors
 to the total flux of the  X-ray background.
 We also  present the colour-colour diagram for our X-ray sources, which
 confirms, together with the spectral analysis on this field performed by
 Mateos et al. (\cite{mateos}) and Mainieri et al. (\cite{mainieri}), the 
 presence of a conspicuous number of highly absorbed AGNs together with evidence of the 
 increasing fraction of this kind of objects at low X-ray fluxes.

\begin{sidewaystable*}
\begin{minipage}[t][180mm]{\textwidth}
\caption[]{Catalogue of X-ray sources with likelihood of
detection in 0.5--10~keV band greater than 10 (3.9 $\sigma$).
Description of columns: (1) XMM-Newton source number as displayed in
Fig. 3, (2) ROSAT source number (Lehmann et al. \cite{lehmann}), (3) IAU source name,
(4--6) J2000 coordinates and errors, (7--18) likelihood of detection (rounded to nearest integer),
fluxes and flux errors [$10^{-16}$ erg cm$^{-2}$ s$^{-1}$] in
0.5--10, 0.5--2, 2--10, and 5--10~keV band, (19--22) hardness ratios and errors
(see text for band definitions). All errors are 1 $\sigma$ errors.}
\label{souCat}
\centering
{\scriptsize
\begin{tabular}{rrcllcrrrrrrrrrrrrrrrr}
\hline
\multicolumn{2}{c|}{SRC \#}&\multicolumn{1}{c|}{}&\multicolumn{2}{c}{RA~~~(J2000)~~~DEC}&err&\multicolumn{3}{|c|}{0.5--10.0~keV}&\multicolumn{3}{|c|}{0.5--2.0~keV}&\multicolumn{3}{|c|}{2.0--10.0~keV}&\multicolumn{3}{|c|}{5.0--10.0~keV}&&&& \\
XMM&\multicolumn{1}{r|}{R}&\multicolumn{1}{c|}{IAU name}&hh mm ss&$^{\circ}$~~~~$^\prime$~~~$^{\prime\prime}$&$^{\prime\prime}$&\multicolumn{1}{|r}{$\mathcal{L}$}&flux&err&\multicolumn{1}{|r}{$\mathcal{L}$}&flux&err&
\multicolumn{1}{|r}{$\mathcal{L}$}&flux&err&\multicolumn{1}{|r}{$\mathcal{L}$}&flux&\multicolumn{1}{r|}{err}&HR1&err&HR2&err \\
\hline
  352 &     &XMMU J105053.0+573028 & 10 50 53.05 & 57 30 28.5 & 2.4 &     13 &    21.0 &  6.1 &     12 &     5.6 &  1.2 &      3 &    15.4 &  6.0 &      0 &     2.6 &  7.6 & -0.39 &  0.19 & -0.79 &  0.56 \\
  358 &  17 &XMMU J105104.0+573053 & 10 51  4.04 & 57 30 53.8 & 0.3 &   2242 &   233.4 &  8.7 &   1896 &    87.1 &  2.3 &    311 &   146.3 &  8.4 &     37 &   111.7 & 15.0 & -0.58 &  0.02 & -0.30 &  0.07 \\
  458 &     &XMMU J105106.4+573435 & 10 51  6.46 & 57 34 35.6 & 0.4 &    744 &   238.9 & 10.7 &    134 &    18.6 &  1.5 &    492 &   220.3 & 10.6 &    118 &   239.6 & 20.0 &  0.30 &  0.04 & -0.13 &  0.05 \\
  430 &  11 &XMMU J105108.3+573345 & 10 51  8.38 & 57 33 45.4 & 0.2 &   3406 &   363.1 & 10.4 &   2522 &   102.0 &  2.4 &    792 &   261.1 & 10.2 &     92 &   186.7 & 16.9 & -0.42 &  0.02 & -0.33 &  0.04 \\
 1522 &     &XMMU J105108.4+572833 & 10 51  8.46 & 57 28 33.6 & 1.7 &     15 &    18.9 &  4.3 &      0 &     0.0 &  0.5 &     11 &    18.9 &  4.3 &      5 &    33.9 & 10.1 &  1.00 &  0.30 &  0.12 &  0.19 \\
  500 &     &XMMU J105108.9+573553 & 10 51  8.97 & 57 35 53.0 & 1.0 &     78 &    62.0 &  8.1 &     40 &    11.1 &  1.4 &     28 &    51.0 &  8.0 &     10 &    79.1 & 18.5 & -0.16 &  0.10 &  0.05 &  0.14 \\
\hline
\end{tabular}}
\vfill
\end{minipage}
\end{sidewaystable*}
\begin{sidewaystable*}
\begin{minipage}[t][180mm]{\textwidth}
\bigskip
\centering
{\scriptsize
\begin{tabular}{rrcllcrrrrrrrrrrrrrrrr}
\hline
\multicolumn{2}{c|}{SRC \#}&\multicolumn{1}{c|}{}&\multicolumn{2}{c}{RA~~~(J2000)~~~DEC}&err&\multicolumn{3}{|c|}{0.5--10.0~keV}&\multicolumn{3}{|c|}{0.5--2.0~keV}&\multicolumn{3}{|c|}{2.0--10.0~keV}&\multicolumn{3}{|c|}{5.0--10.0~keV}&&&& \\
XMM&\multicolumn{1}{r|}{R}&\multicolumn{1}{c|}{IAU name}&hh mm ss&$^{\circ}$~~~~$^\prime$~~~$^{\prime\prime}$&$^{\prime\prime}$&\multicolumn{1}{|r}{$\mathcal{L}$}&flux&err&\multicolumn{1}{|r}{$\mathcal{L}$}&flux&err&
\multicolumn{1}{|r}{$\mathcal{L}$}&flux&err&\multicolumn{1}{|r}{$\mathcal{L}$}&flux&\multicolumn{1}{r|}{err}&HR1&err&HR2&err \\
\hline
 2366 &     &XMMU J105145.3+573215 & 10 51 45.36 & 57 32 15.7 & 1.5 &     13 &    12.9 &  3.0 &      0 &     0.1 &  0.3 &     10 &    12.8 &  3.0 &      4 &    20.4 &  6.8 &  0.93 &  0.26 &  0.06 &  0.20 \\
  426 &     &XMMU J105145.3+573307 & 10 51 45.36 & 57 33  7.2 & 1.3 &     22 &     8.9 &  2.8 &     23 &     4.0 &  0.7 &      1 &     4.8 &  2.8 &      0 &     7.9 &  6.6 & -0.68 &  0.16 &  0.07 &  0.50 \\
  372 &     &XMMU J105146.2+573110 & 10 51 46.25 & 57 31 11.0 & 0.9 &     35 &    20.3 &  3.1 &      0 &     0.6 &  0.5 &     29 &    19.7 &  3.1 &      7 &    21.8 &  6.0 &  0.69 &  0.22 & -0.12 &  0.16 \\
  351 &     &XMMU J105146.5+573036 & 10 51 46.57 & 57 30 36.6 & 0.4 &    462 &    38.4 &  3.2 &    430 &    18.0 &  0.9 &     32 &    20.4 &  3.1 &      2 &    13.3 &  5.5 & -0.70 &  0.04 & -0.37 &  0.19 \\
   16 &     &XMMU J105146.9+571715 & 10 51 46.97 & 57 17 15.0 & 0.5 &    118 &    50.8 &  6.9 &     98 &    16.2 &  1.5 &     17 &    34.6 &  6.7 &      4 &    51.9 & 16.4 & -0.50 &  0.08 &  0.03 &  0.19 \\
 1572 &     &XMMU J105147.4+571840 & 10 51 47.50 & 57 18 40.6 & 1.3 &     14 &    24.8 &  5.2 &      2 &     2.3 &  0.9 &     11 &    22.6 &  5.1 &      1 &    26.1 & 12.8 &  0.22 &  0.22 & -0.10 &  0.27 \\
\hline
\end{tabular}}
\vfill
\end{minipage}
\end{sidewaystable*}
\begin{sidewaystable*}
\begin{minipage}[t][180mm]{\textwidth}
\bigskip
\centering
{\scriptsize
\begin{tabular}{rrcllcrrrrrrrrrrrrrrrr}
\hline
\multicolumn{2}{c|}{SRC \#}&\multicolumn{1}{c|}{}&\multicolumn{2}{c}{RA~~~(J2000)~~~DEC}&err&\multicolumn{3}{|c|}{0.5--10.0~keV}&\multicolumn{3}{|c|}{0.5--2.0~keV}&\multicolumn{3}{|c|}{2.0--10.0~keV}&\multicolumn{3}{|c|}{5.0--10.0~keV}&&&& \\
XMM&\multicolumn{1}{r|}{R}&\multicolumn{1}{c|}{IAU name}&hh mm ss&$^{\circ}$~~~~$^\prime$~~~$^{\prime\prime}$&$^{\prime\prime}$&\multicolumn{1}{|r}{$\mathcal{L}$}&flux&err&\multicolumn{1}{|r}{$\mathcal{L}$}&flux&err&
\multicolumn{1}{|r}{$\mathcal{L}$}&flux&err&\multicolumn{1}{|r}{$\mathcal{L}$}&flux&\multicolumn{1}{r|}{err}&HR1&err&HR2&err \\
\hline
 1591 &     &XMMU J105212.6+572452 & 10 52 12.70 & 57 24 52.6 & 1.3 &     15 &     4.3 &  1.8 &     17 &     2.7 &  0.5 &      0 &     1.6 &  1.8 &      0 &     2.8 &  3.7 & -0.82 &  0.18 &  0.10 &  0.84 \\
  307 &     &XMMU J105212.9+572919 & 10 52 12.99 & 57 29 19.7 & 0.9 &     64 &    21.5 &  2.6 &      0 &     0.2 &  0.3 &     54 &    21.4 &  2.6 &     12 &    20.3 &  4.5 &  0.91 &  0.14 & -0.20 &  0.12 \\
  400 &  13 &XMMU J105213.3+573221 & 10 52 13.35 & 57 32 21.2 & 0.2 &   3273 &   134.4 &  4.5 &   2818 &    52.4 &  1.2 &    423 &    82.0 &  4.3 &     34 &    39.7 &  5.7 & -0.60 &  0.02 & -0.49 &  0.06 \\
  201 &     &XMMU J105213.6+572559 & 10 52 13.63 & 57 25 59.1 & 0.6 &    141 &    24.7 &  2.6 &    103 &     6.4 &  0.6 &     39 &    18.3 &  2.5 &      0 &     6.5 &  4.2 & -0.37 &  0.07 & -0.60 &  0.21 \\
  386 &     &XMMU J105214.0+573140 & 10 52 14.06 & 57 31 40.6 & 0.9 &     46 &    19.0 &  2.7 &      8 &     1.7 &  0.4 &     32 &    17.3 &  2.6 &      6 &    16.1 &  4.3 &  0.23 &  0.14 & -0.21 &  0.15 \\
  588 &     &XMMU J105216.2+574108 & 10 52 16.27 & 57 41  8.6 & 0.4 &    263 &    79.6 &  7.3 &    220 &    25.3 &  1.7 &     44 &    54.3 &  7.1 &      1 &    27.4 & 14.7 & -0.49 &  0.06 & -0.47 &  0.21 \\
\hline
\end{tabular}}
\vfill
\end{minipage}
\end{sidewaystable*}
\begin{sidewaystable*}
\begin{minipage}[t][180mm]{\textwidth}
\bigskip
\centering
{\scriptsize
\begin{tabular}{rrcllcrrrrrrrrrrrrrrrr}
\hline
\multicolumn{2}{c|}{SRC \#}&\multicolumn{1}{c|}{}&\multicolumn{2}{c}{RA~~~(J2000)~~~DEC}&err&\multicolumn{3}{|c|}{0.5--10.0~keV}&\multicolumn{3}{|c|}{0.5--2.0~keV}&\multicolumn{3}{|c|}{2.0--10.0~keV}&\multicolumn{3}{|c|}{5.0--10.0~keV}&&&& \\
XMM&\multicolumn{1}{r|}{R}&\multicolumn{1}{c|}{IAU name}&hh mm ss&$^{\circ}$~~~~$^\prime$~~~$^{\prime\prime}$&$^{\prime\prime}$&\multicolumn{1}{|r}{$\mathcal{L}$}&flux&err&\multicolumn{1}{|r}{$\mathcal{L}$}&flux&err&
\multicolumn{1}{|r}{$\mathcal{L}$}&flux&err&\multicolumn{1}{|r}{$\mathcal{L}$}&flux&\multicolumn{1}{r|}{err}&HR1&err&HR2&err \\
\hline
  443 &     &XMMU J105237.1+573400 & 10 52 37.18 & 57 34  0.4 & 0.3 &     16 &    85.6 &  4.3 &     15 &    15.0 &  0.8 &      2 &    70.6 &  4.2 &      0 &    56.3 &  6.6 & -0.15 &  0.04 & -0.28 &  0.06 \\
  355 &     &XMMU J105237.3+573103 & 10 52 37.35 & 57 31  3.7 & 0.2 &   1561 &    88.2 &  3.7 &   1252 &    30.0 &  0.9 &    287 &    58.2 &  3.6 &     24 &    28.7 &  4.8 & -0.53 &  0.02 & -0.48 &  0.07 \\
  424 &     &XMMU J105238.1+573321 & 10 52 38.12 & 57 33 21.9 & 0.2 &   1859 &   171.2 &  5.5 &    415 &    16.4 &  0.8 &   1153 &   154.8 &  5.4 &    291 &   136.3 &  8.1 &  0.20 &  0.03 & -0.23 &  0.03 \\
 2297 &     &XMMU J105238.5+571644 & 10 52 38.51 & 57 16 44.4 & 1.5 &     22 &    28.9 &  5.5 &     10 &     4.2 &  1.0 &     12 &    24.7 &  5.4 &      1 &    22.4 & 12.6 & -0.04 &  0.16 & -0.22 &  0.29 \\
 1495 &     &XMMU J105239.4+572842 & 10 52 39.49 & 57 28 42.2 & 1.7 &     16 &     7.6 &  1.9 &     12 &     1.9 &  0.4 &      4 &     5.7 &  1.8 &      0 &     5.9 &  3.5 & -0.36 &  0.17 & -0.15 &  0.33 \\
  148 &  32 &XMMU J105239.7+572431 & 10 52 39.76 & 57 24 31.5 & 0.1 & 102046 &  1413.9 & 11.4 &  87917 &   633.0 &  3.6 &  12131 &   780.9 & 10.8 &   1999 &   434.1 & 12.0 & -0.67 &  0.00 & -0.44 &  0.01 \\
\hline
\end{tabular}}
\vfill
\end{minipage}
\end{sidewaystable*}
\begin{sidewaystable*}
\begin{minipage}[t][180mm]{\textwidth}
\bigskip
\centering
{\scriptsize
\begin{tabular}{rrcllcrrrrrrrrrrrrrrrr}
\hline
\multicolumn{2}{c|}{SRC \#}&\multicolumn{1}{c|}{}&\multicolumn{2}{c}{RA~~~(J2000)~~~DEC}&err&\multicolumn{3}{|c|}{0.5--10.0~keV}&\multicolumn{3}{|c|}{0.5--2.0~keV}&\multicolumn{3}{|c|}{2.0--10.0~keV}&\multicolumn{3}{|c|}{5.0--10.0~keV}&&&& \\
XMM&\multicolumn{1}{r|}{R}&\multicolumn{1}{c|}{IAU name}&hh mm ss&$^{\circ}$~~~~$^\prime$~~~$^{\prime\prime}$&$^{\prime\prime}$&\multicolumn{1}{|r}{$\mathcal{L}$}&flux&err&\multicolumn{1}{|r}{$\mathcal{L}$}&flux&err&
\multicolumn{1}{|r}{$\mathcal{L}$}&flux&err&\multicolumn{1}{|r}{$\mathcal{L}$}&flux&\multicolumn{1}{r|}{err}&HR1&err&HR2&err \\
\hline
  403 &     &XMMU J105257.1+573235 & 10 52 57.13 & 57 32 35.6 & 0.5 &    198 &    21.8 &  2.5 &    181 &     8.8 &  0.6 &     19 &    13.0 &  2.4 &      0 &     2.1 &  3.0 & -0.62 &  0.06 & -0.79 &  0.26 \\
  176 &  30 &XMMU J105257.3+572507 & 10 52 57.32 & 57 25  7.3 & 0.1 &   5813 &   180.8 &  4.8 &   5179 &    82.2 &  1.5 &    593 &    98.6 &  4.6 &     43 &    44.6 &  5.9 & -0.68 &  0.01 & -0.52 &  0.05 \\
   19 &     &XMMU J105257.3+571729 & 10 52 57.39 & 57 17 29.7 & 1.2 &     30 &    27.8 &  5.0 &     17 &     4.9 &  0.9 &     13 &    22.9 &  4.9 &      0 &    17.6 & 11.1 & -0.15 &  0.14 & -0.30 &  0.30 \\
 2375 &     &XMMU J105257.3+574152 & 10 52 57.40 & 57 41 52.8 & 0.6 &    186 &    83.7 &  8.1 &    161 &    39.0 &  2.8 &     24 &    44.7 &  7.6 &      2 &    46.6 & 18.8 & -0.69 &  0.05 & -0.15 &  0.21 \\
  569 &     &XMMU J105257.5+573947 & 10 52 57.56 & 57 39 47.8 & 1.2 &     21 &    24.5 &  4.5 &      0 &     1.1 &  0.7 &     18 &    23.5 &  4.5 &      3 &    28.3 & 10.0 &  0.55 &  0.24 & -0.08 &  0.20 \\
  121 & 434 &XMMU J105258.3+572251 & 10 52 58.32 & 57 22 51.0 & 0.3 &    842 &    69.8 &  4.0 &    723 &    25.4 &  1.0 &    116 &    44.4 &  3.9 &      4 &    18.1 &  5.7 & -0.57 &  0.03 & -0.55 &  0.11 \\
\hline
\end{tabular}}
\vfill
\end{minipage}
\end{sidewaystable*}
\begin{sidewaystable*}
\begin{minipage}[t][180mm]{\textwidth}
\bigskip
\centering
{\scriptsize
\begin{tabular}{rrcllcrrrrrrrrrrrrrrrr}
\hline
\multicolumn{2}{c|}{SRC \#}&\multicolumn{1}{c|}{}&\multicolumn{2}{c}{RA~~~(J2000)~~~DEC}&err&\multicolumn{3}{|c|}{0.5--10.0~keV}&\multicolumn{3}{|c|}{0.5--2.0~keV}&\multicolumn{3}{|c|}{2.0--10.0~keV}&\multicolumn{3}{|c|}{5.0--10.0~keV}&&&& \\
XMM&\multicolumn{1}{r|}{R}&\multicolumn{1}{c|}{IAU name}&hh mm ss&$^{\circ}$~~~~$^\prime$~~~$^{\prime\prime}$&$^{\prime\prime}$&\multicolumn{1}{|r}{$\mathcal{L}$}&flux&err&\multicolumn{1}{|r}{$\mathcal{L}$}&flux&err&
\multicolumn{1}{|r}{$\mathcal{L}$}&flux&err&\multicolumn{1}{|r}{$\mathcal{L}$}&flux&\multicolumn{1}{r|}{err}&HR1&err&HR2&err \\
\hline
  475 &   6 &XMMU J105316.8+573550 & 10 53 16.86 & 57 35 50.9 & 0.1 &  91979 &  2154.7 & 17.9 &  72626 &   777.2 &  4.9 &  16338 &  1377.5 & 17.3 &   3016 &   914.3 & 21.6 & -0.56 &  0.00 & -0.36 &  0.01 \\
   66 &     &XMMU J105317.3+572010 & 10 53 17.38 & 57 20 11.0 & 0.8 &    101 &    21.9 &  3.6 &     98 &    11.1 &  1.0 &      5 &    10.7 &  3.4 &      0 &    11.8 &  7.8 & -0.74 &  0.08 & -0.13 &  0.36 \\
  103 &     &XMMU J105317.8+572150 & 10 53 17.89 & 57 21 50.1 & 0.7 &    160 &    29.6 &  3.5 &    140 &    10.9 &  0.9 &     21 &    18.7 &  3.4 &      0 &    10.8 &  6.5 & -0.57 &  0.07 & -0.42 &  0.26 \\
 2353 &     &XMMU J105318.6+572719 & 10 53 18.68 & 57 27 19.6 & 2.3 &     11 &     2.5 &  1.6 &     13 &     2.5 &  0.5 &      0 &     0.0 &  1.5 &      0 &     3.4 &  4.1 & -1.00 &  0.19 &  1.00 &  1.22 \\
   69 &  41 &XMMU J105318.9+572044 & 10 53 18.97 & 57 20 44.7 & 0.3 &    967 &    72.5 &  4.7 &    945 &    45.7 &  1.6 &     25 &    26.9 &  4.4 &      0 &     2.7 &  5.2 & -0.83 &  0.03 & -0.87 &  0.24 \\
   39 &  45 &XMMU J105319.0+571852 & 10 53 19.03 & 57 18 52.9 & 0.2 &   2189 &   181.7 &  7.3 &   1930 &    77.1 &  2.1 &    232 &   104.5 &  7.0 &     29 &    89.3 & 13.8 & -0.65 &  0.02 & -0.25 &  0.08 \\
\hline
\end{tabular}}
\vfill
\end{minipage}
\end{sidewaystable*}
\begin{sidewaystable*}
\begin{minipage}[t][180mm]{\textwidth}
\bigskip
\centering
{\scriptsize
\begin{tabular}{rrcllcrrrrrrrrrrrrrrrr}
\hline
\multicolumn{2}{c|}{SRC \#}&\multicolumn{1}{c|}{}&\multicolumn{2}{c}{RA~~~(J2000)~~~DEC}&err&\multicolumn{3}{|c|}{0.5--10.0~keV}&\multicolumn{3}{|c|}{0.5--2.0~keV}&\multicolumn{3}{|c|}{2.0--10.0~keV}&\multicolumn{3}{|c|}{5.0--10.0~keV}&&&& \\
XMM&\multicolumn{1}{r|}{R}&\multicolumn{1}{c|}{IAU name}&hh mm ss&$^{\circ}$~~~~$^\prime$~~~$^{\prime\prime}$&$^{\prime\prime}$&\multicolumn{1}{|r}{$\mathcal{L}$}&flux&err&\multicolumn{1}{|r}{$\mathcal{L}$}&flux&err&
\multicolumn{1}{|r}{$\mathcal{L}$}&flux&err&\multicolumn{1}{|r}{$\mathcal{L}$}&flux&\multicolumn{1}{r|}{err}&HR1&err&HR2&err \\
\hline
  165 &     &XMMU J105345.0+572419 & 10 53 45.01 & 57 24 19.5 & 1.7 &     13 &    12.1 &  3.2 &      9 &     2.6 &  0.6 &      4 &     9.5 &  3.2 &      0 &     5.8 &  6.3 & -0.27 &  0.19 & -0.40 &  0.48 \\
  370 &     &XMMU J105346.1+573104 & 10 53 46.18 & 57 31  4.4 & 1.1 &     37 &     9.6 &  2.8 &     39 &     5.3 &  0.7 &      0 &     4.3 &  2.7 &      0 &     3.3 &  4.6 & -0.77 &  0.13 & -0.30 &  0.69 \\
 1544 &     &XMMU J105346.2+573608 & 10 53 46.23 & 57 36  8.5 & 1.9 &     12 &     4.4 &  2.1 &     13 &     4.0 &  0.8 &      0 &     0.4 &  1.9 &      1 &    16.3 &  9.5 & -0.97 &  0.14 &  0.93 &  0.31 \\
  461 & 229 &XMMU J105346.4+573510 & 10 53 46.47 & 57 35 10.8 & 0.3 &    932 &    94.3 &  5.9 &    888 &    50.5 &  1.9 &     48 &    43.8 &  5.6 &      0 &     6.0 &  7.3 & -0.76 &  0.03 & -0.83 &  0.20 \\
  200 &     &XMMU J105346.9+572606 & 10 53 46.90 & 57 26  6.2 & 0.3 &    856 &    76.0 &  4.5 &    741 &    29.4 &  1.2 &    105 &    46.6 &  4.3 &     11 &    35.3 &  7.9 & -0.60 &  0.03 & -0.30 &  0.11 \\
  442 & 805 &XMMU J105347.3+573350 & 10 53 47.37 & 57 33 50.0 & 0.4 &    668 &    84.1 &  5.1 &    543 &    27.7 &  1.3 &    122 &    56.4 &  5.0 &      4 &    25.7 &  8.3 & -0.51 &  0.04 & -0.51 &  0.12 \\
\hline
\end{tabular}}
\begin{list}{}{}
\item[$^a$] hardness ratio undefined: fluxes in both constituting bands are zero
\end{list}
\vfill
\end{minipage}
\end{sidewaystable*}
 
\begin{table*}
\caption{Catalogue of extended X-ray sources with likelihood of extent greater than 15.0 (5$\sigma$). 
Description of columns: (1) XMM-Newton source number as displayed in Fig. 3, (2) ROSAT source number (Lehmann et al. \cite{lehmann}),
(3) IAU source name, (4--6) J2000 coordinates and errors, (7--8) likelihood of detection and likelihood of extent (rounded to nearest integer),
(9) source extent (core radius of King profile), (10--11) fluxes and flux errors [$10^{-16}$ erg cm$^{-2}$ s$^{-1}$] 
in 0.5--10~keV band, (12--15) hardness ratios and errors. All errors are 1$\sigma$ errors.}
\label{souCatExt}
\centering
{\scriptsize
\begin{tabular}{rrcllcrrrrrrrrr}
\hline
\multicolumn{2}{c|}{SRC \#}&\multicolumn{1}{c|}{}&\multicolumn{2}{c}{RA~~~(J2000)~~~DEC}&err&\multicolumn{2}{|c|}{$\mathcal{L}$}&\multicolumn{1}{c}{extent}&\multicolumn{2}{|c|}{0.5--10.0~keV}& \\
XMM$^a$&R&\multicolumn{1}{|c|}{IAU name}&hh mm ss&$^{\circ}$~~~~$^\prime$~~~$^{\prime\prime}$&$^{\prime\prime}$&\multicolumn{1}{|r}{detection}&\multicolumn{1}{r|}{extent}&\multicolumn{1}{c|}{``}&flux&err&\multicolumn{1}{|r}{HR1}&err&HR2&err \\
\hline
395   &&XMMU J105151.6+573225& 10 51 51.62 & 57 32 25.0 & 2.3 &  134 &  62 & 12.3 & 127.7 &  40.6 & -0.75 &  0.15 & -0.55 &  0.76 \\
355   &&XMMU J105237.3+573104& 10 52 37.30 & 57 31 04.5 & 0.3 & 1761 &  97 &  1.9 & 122.3 &  46.7 & -0.53 &  0.02 & -0.42 &  0.06 \\
2378  &&XMMU J105242.1+573237& 10 52 42.15 & 57 32 37.0 & 2.2 &   76 &  33 &  6.3 &  25.8 &   6.2 & -0.94 &  0.08 & -1.00 &  2.36 \\
2512- &840&XMMU J105246.1+574044& 10 52 46.18 & 57 40 44.8 & 1.5 &  376 & 134 &  9.7 & 217.4 &  22.8 & -0.80 &  0.05 & -0.56 &  0.46 \\
2515  \\     
476   &827&XMMU J105303.4+573529& 10 53  3.48 & 57 35 29.9 & 0.6 &  431 &  27 &  1.5 &  66.4 &   4.9 & -0.50 &  0.04 & -0.24 &  0.11 \\
 69   & 41&XMMU J105318.8+572045& 10 53 18.87 & 57 20 45.1 & 0.7 & 1560 & 479 &  7.0 & 283.6 &  14.0 & -0.78 &  0.02 & -0.67 &  0.25 \\
2394  &&XMMU J105319.9+573536& 10 53 19.96 & 57 35 36.2 & 5.7 &   38 &  27 & 19.3 & 164.8 &  71.6 & -0.91 &  0.16 & -0.08 &  1.44 \\
409   &&XMMU J105336.8+573257& 10 53 36.89 & 57 32 57.4 & 1.5 &  125 &  41 &  4.1 &  26.2 &   6.0 & -0.91 &  0.08 &  0.55 &  0.38 \\
472   &228&XMMU J105339.9+573522& 10 53 39.94 & 57 35 22.6 & 0.5 & 2408 & 565 &  6.3 & 414.9 &  14.8 & -0.64 &  0.02 & -0.51 &  0.09 \\
144   &131&XMMU J105340.1+572352& 10 53 40.10 & 57 23 52.2 & 0.6 &  565 &  39 &  1.8 &  41.5 &   4.4 & -0.94 &  0.04 &  0.16 &  0.49 \\
 58   &&XMMU J105341.9+575525& 10 53 41.98 & 57 19 36.8 & 0.9 &  235 &  18 &  1.9 &  83.3 &   7.9 & -0.56 &  0.06 & -0.29 &  0.20 \\
461   &229&XMMU J105346.5+573509& 10 53 46.54 & 57 35 09.2 & 0.7 & 1463 & 534 & 12.2 & 759.0 &  34.7 & -0.70 &  0.02 & -0.88 &  0.16 \\
185   &&XMMU J105416.2+572458& 10 54 16.24 & 57 24 58.4 & 1.8 &  131 &  31 &  6.0 & 180.9 &  19.3 & -0.50 &  0.06 & -0.81 &  0.28 \\
\hline
\end{tabular}}
\begin{list}{}{}
\item[$^a$]source number of corresponding object from point source detection list (Table~\ref{souCat}), except for object 2394 not previously detected as a point source.
\end{list}
\end{table*}

\begin{acknowledgements}
    In Germany the XMM-{\em Newton} project is supported by the Bundesministerium f\"ur Bildung und Forschung/Deutsches
    Zentrum f\"ur Luft und Raumfahrt and the Max Planck Society. Part of this work was supported by the DLR project
    numbers 50 OR 0207 and 50 OR 0405.  XB acknowledges support from the Spanish Ministry of Education and Science through project ESP2006-13608-C02-01.
\end{acknowledgements}

\end{document}